\documentclass[12pt]{article}

\usepackage{amssymb,amsthm,amscd,array,stmaryrd,xypic}

\let\ssection=\section
\renewcommand{\section}{\setcounter{equation}{0}\ssection}

\newcommand{\bbR}{\mathbb{R}}

\newcommand{\bbC}{\mathbb{C}}

\newcommand{\bbZ}{\mathbb{Z}}

\newcommand{\rA}{\mathrm{A}}

\newcommand{\Aff}{\mathrm{Aff}}

\newcommand{\AO}{\mathrm{AO}}
\newcommand{\cA}{{\mathcal{A}}}
\newcommand{\cC}{{\mathcal{C}}}

\newcommand{\Cinfty}{{\mathcal{C}^{\infty}}}

\newcommand{\cE}{{\mathcal{E}}}
\newcommand{\Diff}{\mathrm{Diff}}

\newcommand{\rE}{\mathrm{E}}
\newcommand{\EO}{\mathrm{EO}}

\newcommand{\cF}{{\mathcal{F}}}

\newcommand{\GL}{{\mathrm{GL}}}

\newcommand{\cM}{{\mathcal{M}}}
\newcommand{\Id}{\mathrm{Id}}

\newcommand{\ord}{\mathrm{ord}}

\newcommand{\PC}{\mathrm{PC}}

\newcommand{\cQ}{{\mathcal{Q}}}

\newcommand{\rO}{\mathrm{O}}
\newcommand{\rP}{\mathrm{P}}

\newcommand{\cS}{{\mathcal{S}}}
\newcommand{\rS}{\mathrm{S}}

\newcommand{\SII}{\mathrm{S}^{1\vert1}}

\newcommand{\SL}{\mathrm{SL}}
\newcommand{\PGL}{\mathrm{PGL}}
\newcommand{\PSL}{\mathrm{PSL}}

\newcommand{\Sp}{\mathrm{Sp}}
\newcommand{\SpO}{\mathrm{SpO}}

\newcommand{\ve}{\varepsilon}
\newcommand{\Vect}{\mathrm{Vect}}

\newcommand{\half}{\frac{1}{2}}

\newcommand{\la}{\langle}
\newcommand{\ra}{\rangle}

\def\a{\alpha}
\def\b{\beta}
\def\g{\gamma}
\def\d{\delta}

\def\l{\lambda}

\newtheorem{thm}{Theorem}[section]
\newtheorem{lem}[thm]{Lemma}
\newtheorem{cor}[thm]{Corollary}
\newtheorem{pro}[thm]{Proposition}

\newtheorem{rmk}[thm]{Remark}
\newtheorem{defi}[thm]{Definition}
\newcommand{\KEYWORDS}[1]{\textbf{Keywords}: #1}

\newcommand{\PREPRINT}[1]{\textbf{Preprint}: #1}

\title{On the projective geometry of the supercircle: a unified construction of the super cross-ratio and Schwarzian derivative}
\author{
{\sc J.-P.~Michel}\footnote{mailto: michel@cpt.univ-mrs.fr}
\and
{\sc C.~Duval}\footnote{mailto: duval@cpt.univ-mrs.fr}\\\\
Centre de Physique Th\'eorique, CNRS, 
Luminy, Case 907\\ 
F-13288 Marseille Cedex 9 (France)\footnote{ 
UMR 6207 du CNRS associ\'ee aux 
Universit\'es d'Aix-Marseille I et II et Universit\'e du Sud Toulon-Var; Laboratoire 
affili\'e \`a la FRUMAM-FR2291
}
}

\date{}

\begin{document}

\maketitle
\thispagestyle{empty}

\begin{abstract}
We consider the standard contact structure on the supercircle, $\rS^{1|1}$, and the supergroups $\rE(1|1)$, $\Aff(1|1)$ and $\SpO(2|1)$ of contactomorphisms, defining the Euclidean, affine and projective geometry respectively. Using the new notion of $p|q$-transitivity, we construct in synthetic fashion even and odd invariants characterizing each geometry, and obtain an even and an odd super cross-ratios.

Starting from the even invariants, we derive, using a superized Cartan formula, one-cocycles of the group of contactomorphisms, $K(1)$, with values in tensor densities $\cF_\l(\rS^{1|1})$. The even cross-ratio yields a $K(1)$ one-cocycle with values in quadratic differentials, $\cQ(\rS^{1|1})$, whose projection on $\cF_\frac{3}{2}(\rS^{1|1})$ corresponds to the super Schwarzian derivative arising in superconformal field theory. This leads to the classification of the cohomology spaces $H^1(K(1),\cF_\l(\rS^{1|1}))$. 

The construction is extended to the case of $\rS^{1|N}$. All previous invariants admit a prolongation for $N>1$, as well as the associated Euclidean and affine cocycles. The super Schwarzian derivative is obtained from the even cross-ratio, for $N=2$, as a projection to $\cF_1(\rS^{1|2})$ of a $K(2)$ one-cocycle with values in~$\cQ(\rS^{1|2})$. The obstruction to obtain, for $N\geq 3$, a projective cocycle is pointed out.
\end{abstract}

%\bigskip

\KEYWORDS{Projective structures, Contact geometry, Supermanifolds, Cross-ratio, Schwarzian derivative}

%\medskip

%\AMSCLASSIFICATION{78A05, 70G45, 58B20}

%\medskip

\PREPRINT{CPT-P37-2007}

%\newpage

\tableofcontents

\baselineskip=19pt

%%%%%%%%%%%%%%%%%%%%%%%%%%%%%%%%%%%%%%%%%%%%%%%%%%%%%%%%%%%%
%%%%%%%%%%%%%%%%%%%%%%%%%%%%%%%%%%%%%%%%%%%%%%%%%%%%%%%%%%%%
\section{Introduction}\label{Intro}
%%%%%%%%%%%%%%%%%%%%%%%%%%%%%%%%%%%%%%%%%%%%%%%%%%%%%%%%%%%%
%%%%%%%%%%%%%%%%%%%%%%%%%%%%%%%%%%%%%%%%%%%%%%%%%%%%%%%%%%%%

The cross-ratio is the fundamental object of projective geometry; it is
 a projective invariant of the circle $S^1$ (or, rather, of 
$\bbR\rP^1$). The main objective of this article is to propose and
 justify from a group theoretical analysis a super-analogue of the cross-ratio
 in the case of the super\-circle~$\rS^{1|N}$, and to deduce then, from the
 Cartan formula (\ref{TheCartanFormula}), the associated Schwarzian derivative for $N=1,2$.

It is well-known that the circle, $S^1$, admits three different
 geometries, namely the Euclidean, affine and projective geometries, as highlighted by Ghys~\cite{Ghy}. They are
 defined by the groups $(\bbR,+)$, $\Aff(1,\bbR)$ and $\PGL(2,\bbR)$, or
 equivalently by their characteristic invariants, the distance, the
 distance ratio, and the cross-ratio. From these invariants we can obtain,
 using Cartan-like formul{\ae}, three $1$-cocycles of
 $\Diff_+(\rS^1)$ with coefficients in some tensorial density modules
 $\cF_\l(\rS^1)$ with $\l\in\bbR$; see \cite{DG}. They are the generators of the three nontrivial cohomology spaces $H^1(\Diff_+(\rS^1),\cF_\l)$, with
 $\l=0,1,2$, as proved in \cite{Fuk}. 

The purpose of this article is to extend these results to the
 supercircle, $\rS^{1|N}$, endowed with its standard contact structure. To that end, we use the
 embedding of the quotient, $\PC(2|N)=\SpO(2|N)/\{\pm\Id\}$, of the orthosymplectic supergroup $\SpO(2|N)$, into the group, $K(N)$, of contactomorphisms of $\rS^{1|N}$. The supergroup $\PC(2|N)$ is the projective conformal supergroup introduced by Manin in \cite{Man}, extending $\PSL(2,\bbR)$.
The two main objects of super projective geometry, namely the
 cross-ratio and the Schwarzian derivative, have, indeed, already been introduced in the
 general context of superstring theories, though in a somewhat
 independent fashion. This was mainly done in the framework of super Riemann
 surfaces, or in terms of the so-called SUSY structures. On the one hand, the
 even and odd cross-ratios, for $N=1$, have been originally put forward
 by Aoki \cite{Aok}, and Nelson \cite{Nel}, respectively; these two
 references have opened the way to subsequent work of, e.g., Giddings
 \cite{Gid}, and Uehara and Yasui~\cite{UY}. On the other hand, the super
 Schwarzian derivative has been introduced, in the framework of
 superconformal field theory, by Friedan \cite{Fri}, for $N=1$, and by Cohn~\cite{Coh}, for $N=2$. 
 
 \goodbreak
 
Quite independently, and from a more mathematical point of view, Manin
 \cite{Man} introduced the even and odd cross-ratios, for $N=1,2$, by
 resorting to linear super\-symplectic algebra. Also did Radul
 \cite{Rad0,Rad} discover the formul{\ae} for the super Schwarzian $K(N)$
 $1$-cocycles, for $N=1,2,3$, using the transformation laws of the super
 Sturm-Liouville operators on $\rS^{1|N}$.

Our first objective is to construct, in a systematic manner, invariants characterizing each supergroup $\rE_+(1|N)\subset\Aff_+(1|N)\subset\PC(2|N)$ acting on the supercircle~$\rS^{1|N}$. To this end, we introduce the new notion of $p|q$-transitivity, well-adapted to supergroups, and state a general theorem, providing a way to build up characteristic invariant of a simply $p|q$-transitive group action. Applying this theorem to the three preceding supergroups, we obtain Euclidean, affine, and projective invariants, respectively $I_\mathsf{e}$, $I_\mathsf{a}$ and~$I_\mathsf{p}$, with their even and odd part. In the case $N=1$, the two components of $I_\mathsf{p}$ are, unsurprisingly,
 the even and odd above-mentioned super cross-ratios. Let us emphasize that,
 for arbitrary $N$, the even cross-ratio turns out to be given by the
 superfunction
\begin{equation}
[t_1,t_2,t_3,t_4]=\frac{[t_1,t_3][t_2,t_4]}{[t_2,t_3][t_1,t_4]}   
\label{TheSuperCrossratio}  
\end{equation}
of a quadruple of ``points'' $(t_1,t_2,t_3,t_4)$ of $\rS^{1|N}$, with even
 coordinates $x_i$, and odd ones $(\xi_i^1,\ldots,\xi_i^N)$, for
 $i=1,\ldots,4$; note that in (\ref{TheSuperCrossratio}) the two-point superfunction
$[t_i,t_j]=x_j-x_i-\xi_j\cdot\xi_i$ is the Euclidean even invariant. The supergroups preserving $I_\mathsf{e}$, $I_\mathsf{a}$ and $I_\mathsf{p}$ are respectively $\rE_+(1|N)$, $\Aff_+(1|N)$ and $\PC(2|N)$, as expected.

Our second objective, is to link the three even parts of the previously found invariants to $1$-cocycles of $K(N)$, by means of a natural superized version of the Cartan formula. It culminates in the projective case, where we get the super Schwarzian derivative (\ref{Projectivecocycle}) from the even cross-ratio. Let us go into some more details. Given a flow, $\phi_\ve=\Id+\ve X+O(\ve^2)$, we posit
 $t_{i+1}=\phi_{i\ve}(t_1)$, for $i=0,\ldots,3$. We contend that the Cartan formula
 \cite{Car,OT} can be consistently superized for $N=1$, and $N=2$, using the
 cross-ratio (\ref{TheSuperCrossratio}), namely by
\begin{equation}
\frac{\Phi^*[t_1,t_2,t_3,t_4]}{[t_1,t_2,t_3,t_4]}-1=\left\langle \ve 
X\otimes \ve X, \cS(\Phi)\right\rangle + O(\ve^3),
\label{TheCartanFormula}
\end{equation}
hence, providing us with a definition of the Schwarzian derivative,
 $\cS(\Phi)$, of a contacto\-morphism $\Phi$. In doing so, we naturally
 obtain a  $1$-cocycle of $K(N)$, for $N=1,2$ respectively, with
 values in the module, $\cQ(\rS^{1|N})$, of quadratic differentials. Their projections onto
 the modules $\cF_{\frac{3}{2}}(\SII)$ and $\cF_1(\rS^{1|2})$, for $K(1)$ and $K(2)$
 respectively, yield the expressions of the super Schwarzian derivatives given in \cite{Fri,Coh,Rad}.
 Remarkably enough, our formula  allows us to recover the classical
 Schwarzian derivative on the circle, $\rS^1$, which would not be the case, had we
 started with Friedan's, Cohn's, and Radul's formul{\ae}. Much in the same way, we define the Euclidean and affine $1$-cocycles of $K(N)$ for any $N$, with the help of the Cartan-like formul{\ae} (\ref{CartanEuclid}), and (\ref{CartanAffine}). 
 Using the results of Agrebaoui et
 al. \cite{ABF} on the cohomology of the Lie superalgebra of contact
 vector field on $\SII$, we can claim that our three $1$-cocycles on $K(1)$
 are, indeed, the generators of the three nontrivial cohomology
 spaces $H^1(K(1),\cF_\l)$, where $\l=0,\half,\frac{3}{2}$. 
 
\medskip
\goodbreak

The paper is organized as follows.

In Section \ref{S11}, we recall the main definitions and facts related
 to the geometry of the supercircle $\rS^{1|1}$, in particular its
 canonical contact structure and the action of the (special) orthosymplectic
 group $\SpO_+(2|1)\cong{}\PC(2|1)$, as a subgroup of the group, $K(1)$, of
 contactomorphisms of $\rS^{1|1}$.

In Section \ref{MainResults}, we review the main results of this
 article, namely the form of the invariants, and of the associated
 $1$-cocycles of $K(1)$, obtained for each of the three above-mentioned
 geometries. This section also gives the classification of the cohomo\-logy
 spaces $H^1(K(1),\cF_\l)$, for $\l\in\bbC$.

Sections \ref{InvariantSection} and \ref{CartanSection} provide the
 proofs of the main results announced in Section~\ref{MainResults}. We first define the notion of $p|q$-transitivity and state the general Theorem \ref{generalinvariant}, leading to the
 construction of the Euclidean, affine, and projective invariants, from the action of the corresponding subgroups of
 $K(1)$. Those invariants are then shown to yield, via a Taylor expansion, the
 sought $1$-cocycles; in particular the Cartan formula readily leads to a
 new expression for the Schwarzian derivative, $\cS(\Phi)$, of a
 contactomorphism, $\Phi$, with values in the module of quadratic
 differentials, $\cQ(\SII)$. The link with Friedan's and Radul's Schwarzian
 derivative is elucidated. The kernels of the three above $1$-cocycles are shown
 to be, indeed, isomorphic to $\rE(1|1)$, $\Aff(1|1)$, and $\SpO_+(2|1)$
 respectively.
 
 \goodbreak

In Section \ref{S1N}, we present a detailed treatment of the general
 case, $N>1$, along the same lines as before. As mentioned in Section
 \ref{MainResults}, there is hardly no change in the construction and the
 resulting expressions of the invariants. The Euclidean and affine
 $1$-cocycles of $K(N)$ are explicitly derived, as well as the Schwarzian
 derivative obtained as a $1$-cocycle of $K(2)$ with values in the module
 $\cQ(\rS^{1|2})$ of quadratic differentials. Upon projection of $\cQ(\rS^{1|2})$ onto the $K(2)$-module $\cF_1(\rS^{1|2})$ of $1$-densities, we obtain Cohn's and Radul's formula for the Schwarzian derivative. Specific difficulties encountered in deriving the projective
 $1$-cocycles for $N>2$ are pointed out, together with those arising in
 the determination of the kernels of the Euclidean and affine
 $1$-cocycles. At last, the kernel of the Schwarzian $1$-cocycle of $K(2)$ is
 shown to be isomorphic to~$\PC(2|2)$.

Section \ref{Conclusion} gives us the opportunity to sum up the content
 of this article, and to draw several conclusions. It opens
 perspectives for future work related to the link between discrete projective
 invariants of the supercircle, and the cohomology of the group of its
 contactomorphisms.  

 \subsection*{Acknowledgements}
 It is a pleasure to acknowledge most enlightening discussions with V.~Fock, D. Leites, and C.~Roger. Special thanks are due to V.~Ovsienko for his constant interest in this work and a number of suggestions that have greatly improved this paper.

%%%%%%%%%%%%%%%%%%%%%%%%%%%%%%%%%%%%%%%%%%%%%%%%%%%%%%%%%%%%
%%%%%%%%%%%%%%%%%%%%%%%%%%%%%%%%%%%%%%%%%%%%%%%%%%%%%%%%%%%%
\section{The supercircle $\SII$ and its contacto\-morphisms: A 
compendium}\label{S11}
%%%%%%%%%%%%%%%%%%%%%%%%%%%%%%%%%%%%%%%%%%%%%%%%%%%%%%%%%%%%
%%%%%%%%%%%%%%%%%%%%%%%%%%%%%%%%%%%%%%%%%%%%%%%%%%%%%%%%%%%%
We briefly define in this section the geometrical objects on $\SII$ 
that will be needed for our purpose.
This includes the basics of super differential geometry \cite{Lei,Man0,DM}, 
the standard contact structure on the supercircle \cite{Rad}, and 
the orthosymplectic group $\SpO(2|1)$, see~\cite{Man}.  

\goodbreak

%%%%%%%%%%%%%%%%%%%%%%%%%%%%%%%%%%%%%%%%%%%%%%%%%%%%%%%%%%%%
\subsection{The supercircle $\SII$}  
%%%%%%%%%%%%%%%%%%%%%%%%%%%%%%%%%%%%%%%%%%%%%%%%%%%%%%%%%%%%

The supercircle $\SII$ can be defined as the circle, $\rS^1$, endowed with the
 sheaf of the super\-commutative associative algebra of superfunctions
 $\Cinfty(\SII)=\Cinfty(\rS^1)[\xi]$. Thus, $\SII$ admits local coordinates $t=(x,\xi)$, where $x$ is a local coordinate on $S^1$, and $\xi$ is an odd (Grassmann) coordinate, i.e., such that
 $\xi^2=0$ and $x\xi=\xi x$. Then, a superfunction is of the form
\begin{equation}
f(x,\xi)=f_0(x)+\xi f_1(x)
\label{superfunction}
\end{equation} 
with $f_0,f_1\in\Cinfty(\rS^1)$. There exists a 
$\bbZ_2$-grading on superfunctions, $f_0$ being the even part and $\xi
 f_1$ the odd part of $f$. The parity is denoted by $p$, with the convention 
$p(f_0)=0$ and $p(\xi f_1)=1$. We define the projection
\begin{equation} \label{quotient}
\pi:\Cinfty(\SII)\rightarrow\Cinfty(\rS^1)
\end{equation}
by quotienting by the ideal of nilpotent elements; this gives an
 embedding of the circle into the supercircle.
 
 \goodbreak

Denote by $\Diff(\SII)$ the group of diffeomorphisms of $\SII$,
 i.e., the group of automorphisms of $\Cinfty(\SII)$. Let 
$\Phi\in\Diff(\SII)$, then
\begin{equation}
\Phi(x,\xi)=(\varphi(x,\xi),\psi(x,\xi))
\label{superdiffeomorphism}
\end{equation}   
where $\varphi$ is an even superfunction and $\psi$ an odd one, so 
$\Phi$ preserves parity and $(\varphi(x,\xi),\psi(x,\xi))$ become new 
coordinates on $\SII$. For any morphism, i.e., algebra morphism
 preserving parity, the following diagram is commutative
\begin{equation}\label{diagram}
\xymatrix{
\Cinfty(\SII)  \ar[r]^{\pi} 
& \Cinfty(\rS^1) \\
\Cinfty(\SII) \ar[u]^{\Phi} \ar[r]^{\pi}
& \Cinfty(\rS^1) \ar[u]_{\Pi(\Phi)} 
}
\end{equation}
So, every morphism of $\Cinfty(\SII)$ induces a morphism of
 $\Cinfty(\rS^1)$, and we have a canonical morphism $\Pi:\Diff(\SII)\to\Diff(\rS^1)$.

A super vector field, $X$, on $\SII$ is a superderivation of 
$\Cinfty(\SII)$, i.e., a linear operator satisfying super Leibniz rule, $X(fg)=X(f)g+(-1)^{p(f)p(X)}fX(g)$, for homogeneous elements. As in ordinary differential geometry, $X$ can be locally written in terms of partial derivatives as
\begin{equation}
X=f(x,\xi)\partial_x+g(x,\xi)\partial_\xi
\label{supervectorfield}
\end{equation} 
where $f,g\in\Cinfty(\SII)$, with $p(\partial_x)=0$ and 
$p(\partial_\xi)=1$. The space, $\Vect(\SII)$, of vector fields on
 $\SII$ 
is thus a left-module over $\Cinfty(\SII)$. It has the structure of a
 super Lie algebra, $\Vect(\SII)=\Vect(\SII)_0\oplus\Vect(\SII)_1$, whose superbracket is denoted by $[\,\cdot\,,\,\cdot\,]$, and $[X,Y]=XY-(-1)^{p(X)p(Y)}YX$, for homogeneous elements.

Since the group $\Diff(\SII)$ of diffeomorphisms preserves parity, we
 can define the 
flow of $X\in\Vect(\SII)$, namely $\varphi_\ve=\Id+\ve X +\rO(\ve^2)$, 
only if $p(\ve X)=0$. For odd vector fields, $X$, the parameter $\ve$
 must therefore be odd, see 
\cite{DM}. 

We can now define the $\Cinfty(\SII)$ right-module $\Omega^1(\SII)$ of 
$1$-forms on $\SII$, as the dual of the $\Cinfty(\SII)$ left-module 
$\Vect(\SII)$. The $1$-forms $dx$ and $d\xi$ will constitute the dual 
basis of $\partial_x$ and $\partial_\xi$, that is $\la \partial_x , 
dx\ra=\la \partial_\xi , d\xi\ra=1$ and $\la \partial_\xi , dx\ra=\la 
\partial_x , d\xi\ra=0$. Then $p$ is extended naturally to
 $\Omega^1(\SII)$ by 
$p(dx)=0$ and $p(d\xi)=1$. 
Using the exterior product we construct $\Omega^*(\SII)$, the space of
 all 
differential forms on $\SII$, graded by $\bbZ$ with $\vert\cdot\vert$
 the 
cohomological degree. Parity being also defined on this space, we have
 two choices for the Sign Rule, viz.,
\begin{eqnarray}
\a\wedge\b&=&(-1)^{(p(\a)+|\a|)\,(p(\b)+|\b|)}\b\wedge\a \\
\a\wedge\b&=&(-1)^{|\a||\b|+p(\a)p(\b)}\b\wedge\a
\label{signrule}
\end{eqnarray} 
where $\a,\b$ are homogeneous elements of $\Omega^*(\SII)$. The second
 convention corresponds to a bigrading $\bbZ\times\bbZ_2$, and, 
following \cite{DM, Kostant}, we will choose it from now on.

%%%%%%%%%%%%%%%%%%%%%%%%%%%%%%%%%%%%%%%%%%%%%%%%%%%%%%%%%%%%
\subsection{The contact structure on $\SII$ and its automorphisms}
\label{ContactS11}
%%%%%%%%%%%%%%%%%%%%%%%%%%%%%%%%%%%%%%%%%%%%%%%%%%%%%%%%%%%%

The standard contact structure on $\SII$ is given by the conformal 
class of the $1$-form
\begin{equation}
\a=dx+\xi d\xi
\label{contactform}
\end{equation}
which satisfies $\alpha\wedge{}d\alpha\neq0$.  This contact structure 
is equivalently defined by the kernel of~$\a$, spanned by the odd
 vector field 
\begin{equation}\label{D}
D=\partial_\xi+\xi\partial_x,
\end{equation}
whose square $D^2=\half[D,D]=\partial_x$ is the Reeb vector field of
 the structure.

\goodbreak

Then $D$ and $\partial_x$ set up a basis of the $\Cinfty(\SII)$ 
left-module $\Vect(\SII)$, while $\a$ and $\b=d\xi$ constitute the dual
 basis, with 
$d\a=\b\wedge\b$. Thus for any $f\in\Cinfty(\SII)$ we have
\begin{equation}
df=\a f'+\b Df.
\label{df}
\end{equation} 
where $f'=\partial_x f$. The contact structure being given by the 
direction of $\a$, it is therefore preserved by $\Phi\in\Diff(\SII)$
 iff 
\begin{equation}
\Phi^*\a=E_\Phi\a
\label{contactomorphism}
\end{equation} 
for some superfunction $E_\Phi$, which, following \cite{Rad}, we call the multiplier of
 $\Phi$. We denote by $K(1)$ the subgroup of $\Diff(\SII)$ preserving the contact
structure, its elements are called contactomorphisms.
From (\ref{superdiffeomorphism}) and (\ref{contactform}) we find
 $\Phi^*\a=d\varphi + \psi 
d\psi= \a (\varphi'+\psi \psi')+ \b(D\varphi-\psi D\psi)$. 

\begin{pro}\label{lemcontactomorphism}
Let $\Phi=(\varphi,\psi)$ be a diffeomorphism of $\SII$; then $\Phi\in 
K(1)$ iff 
\begin{equation}
D\varphi-\psi D\psi=0.
\label{contactconstraint}
\end{equation}   
The multiplier of $\Phi$ is then given by $E_\Phi=\varphi'+\psi\psi'$,
 i.e., by
\begin{equation}
E_\Phi=\frac{\Phi^*\a}{\a}=(D\psi)^2.
\label{E}
\end{equation}
\end{pro}

\goodbreak

Since $\a$ and $\b$ set up a basis of the $\Cinfty(\SII)$-module 
$\Omega^1(\Cinfty(\SII))$, we will also need the expression of the action of $K(1)$ on the
 odd $1$-form $\b$; it reads
\begin{equation}
\Phi^*\b=\a \psi'+\b D\psi.
\label{btransformation}
\end{equation} 
We might, as well, define $K(1)$ as the group of diffeomorphisms 
preserving the horizontal distribution spanned by $D$, denoted by $\la D\ra$. In the complex setting, $D$ is interpreted as the covariant derivative of a super Riemann surface \cite{Fri,Coh}, and $K(1)$ as the superconformal group; the distribution $\la D\ra$ is also often referred to as a SUSY structure \cite{Man,DM}. See also \cite{GR} for a review. 

Using (\ref{contactomorphism}), we find that the transformation law (\ref{btransformation}) entails
\begin{equation}
\Phi^*D=\frac{1}{D\psi} D,
\label{Dtransformation}
\end{equation}
which makes sense as $D\psi\neq0$ for any diffeomorphism $\Phi$.

\begin{rmk}\label{orientation}
{\rm
If $\Phi=(\varphi,\psi)\in{}K(1)$, see (\ref{superdiffeomorphism}), we
 put $\varphi(x,\xi)=\varphi_0(x)+\xi\varphi_1(x)$, and $\psi(x,\xi)=
 \psi_1(x)+\xi \psi_0(x)$, with an index $0$ for even functions and $1$ for odd
 functions. The constraint (\ref{contactconstraint}) then reads
 $\varphi'_0=\psi_0^2-\psi_1\psi'_1$ and $\varphi_1=\psi_0\psi_1$. Using the
 natural projection $\Pi:K(1)\to\Diff(S^1)$, defined in (\ref{diagram}), we note that
 $\Phi$ gives rise to a diffeomorphism of $S^1$, which is actually
 orientation-preserving since $\Pi(\Phi)'=\pi(\varphi_0')=\pi(\psi_0)^2>0$.
}
\end{rmk}

From the constraint (\ref{contactconstraint}), we can obtain an
 interesting property of contactomorphisms: they are essentially determined by
 their even part.

\begin{lem}\label{even-all}
Let $\Phi=(\varphi,\psi)\in K(1)$ and
 $\widetilde{\Phi}=(\widetilde{\varphi},\widetilde{\psi})\in K(1)$, be two contacto\-morphisms such that
 their even part coincide, $\varphi=\widetilde{\varphi}$. We then have
 $\widetilde{\psi}=\pm\psi$.
\end{lem}
This can be checked by a direct calculation.

%-----------------------------------------------------------
\subsubsection{The super Lie algebra, $k(1)$, of contact vector fields}
%-----------------------------------------------------------

In view of the definition (\ref{contactomorphism}) of 
contactomorphisms, we will call $X\in\Vect(\SII)$ a contact vector
 field, $X\in k(1)$, 
if
\begin{equation}
L_X\a=e_X\,\a 
\end{equation}
for some superfunction $e_X$.
The Lie derivative is still given by the derivative of the flow, so 
$k(1)$ is the Lie algebra of $K(1)$, and $e$ is the 
derivative of $E$ at the identity. Let us now recall the following classic
result \cite{GLS,GMO}: if $X\in k(1)$, there exists a unique superfunction 
$f(x,\xi)=a(x)-2\xi b(x)$, called the contact Hamiltonian, such that~$X=X_f$, where
\begin{equation}
X_f=a(x)\partial_x+\frac{1}{2}a'(x)\xi\partial_\xi+b(x)(\partial_\xi-\xi\partial_x)
\label{contactvectorfield}
\end{equation} 
so that the associated (infinitesimal) multiplier is given by
\begin{equation}
e_{X_f}=f'.
\label{e(Xf)}
\end{equation}

%-----------------------------------------------------------
\subsubsection{Tensor densities, $1$-forms and quadratic differentials of $\SII$}
\label{omega,q}
%-----------------------------------------------------------

Let us introduce now a $1$-parameter family,  $\cF_\lambda(\SII)$ 
or $\cF_\lambda$ for short, of $K(1)$-modules, which define the 
$\l$-densities associated with the contact structure, $\l\in\bbC$. As
 vector spaces, these 
modules are isomorphic to $\Cinfty(\SII)$, the $K(1)$ anti-action 
$(\Phi\mapsto\Phi_{\l})$ on $\cF_\lambda(\SII)$ being given by
\begin{equation}
\Phi_\lambda f= (E_\Phi)^{\lambda}\;\Phi^*f,
\label{densitytransformation}
\end{equation}
where $f\in\Cinfty(\SII)$. 
We may thus write a $\l$-density $F\in\cF_\lambda$,  
symbolically, as~$F=f\a^\l$. We will thus write $(\Phi\to\Phi^*)$
 the $K(1)$ anti-action on $\cF_\l$ with this identification. 
\begin{rmk}\label{Ddensity}
{\rm
In view of (\ref{E}) and (\ref{Dtransformation}), we will regard, in
 conformity with the definition (\ref{densitytransformation}), the odd
 vector field $D$ as a $(-\half)$-density.
}
\end{rmk}

There is an isomorphism of $K(1)$-modules: $\Vect(\SII)\cong 
\cF_{-1}\oplus\cF_{-\half}$, where $\cF_{-1}$ corresponds to $k(1)$ and
 $\cF_{-\half}$ 
to the vector fields $fD$, with $f\in\Cinfty(\SII)$ and $D$ as in~(\ref{D}). See \cite{GLS,GMO}. The space of $1$-forms $\Omega^1(\SII)$ is
 generated, as 
$\cC^\infty(\SII)$-module, by $\a$ and $\b$. 

Similarly the space
 $\cQ(\SII)$ of quadratic differentials is generated, as a
 $\cC^\infty(\SII)$-module by
\begin{equation}
\a^2=\a\otimes\a \quad 
\mbox{ and } 
\quad 
\a\b=\half(\a\otimes\b+\b\otimes\a),
\label{tensorproduct}
\end{equation}
where the tensor product is understood as the supersymmetric tensor
 product constructed via the commutativity isomorphism given by the Sign
 Rule \cite{DM}. This notation will be used throughout this paper. 
\begin{pro}\label{split}
The two $K(1)$-modules $\Omega^1(\SII)$ and $\cQ(\SII)$, admit the
 following decomposition into $K(1)$-submodules, namely
\begin{eqnarray}
\Omega^1(\SII)&\cong&\cF_\half\oplus\cF_1, \\
\cQ(\SII)&\cong&\cF_\frac{3}{2}\oplus\cF_2. 
\end{eqnarray}
The summands $\cF_1$ (resp. $\cF_2$) are naturally $K(1)$-submodules of
 $\Omega^1(\SII)$ (resp. $\cQ(\SII)$). The projections
 $\Omega^1(\SII)\to\cF_\half$ (resp. $\cQ(\SII)\to\cF_\frac{3}{2}$) are given by
 $\a^\half\la{}D,\,\cdot\,\ra$, and the corresponding sections by $\a^\half
 L_D$ (resp. $\frac{2}{3}\a^\half L_D$).
\end{pro}

\begin{proof}
We have $\a^\half\la{}D,\a f+\b g\ra=\a^\half g$ and
 $\a^\half\la{}D,\a^2 f+\a\b g\ra=\half\a^\frac{3}{2} g$. The transformation rules (\ref{E}) for $\a$ and (\ref{btransformation}) for $\b$ then entail that the
 projections $\a^\half\la{}D,\,\cdot\,\ra$ actually define morphisms of
 $K(1)$-modules, $\Phi^*(\a^\half\la D, \omega\ra)=\a^\half \la D,
 \Phi^*\omega\ra$ for all $\omega\in \Omega^1(\SII)$, and for all $\omega\in
 \cQ(\SII)$. 

Moreover, since $L_D\a=2\b$ and $L_D\b=0$, we readily find $\a^\half
 L_D(\a^\half g)=\a Dg +\b g$ and $\a^\half L_D(\a^\frac{3}{2} g)=\a^2 Dg
 +3\a\b g$. Using, once more, (\ref{E}) and (\ref{btransformation}),
 we then obtain that the inclusions $\a^\half L_D$ define, again,
 morphisms of $K(1)$-modules. To have the identity $\mu\a^\half\la{}D,\a^\half L_D
 F\ra=F$, we choose $\mu=1$ for $F$ a $\half$-density, and $\mu=\frac{2}{3}$ for $F$ a $\frac{3}{2}$-density. The result follows.
\end{proof}

%%%%%%%%%%%%%%%%%%%%%%%%%%%%%%%%%%%%%%%%%%%%%%%%%%%%%%%%%%%%
\subsection{The orthosymplectic group $\SpO(2|1)$}\label{SpO}
%%%%%%%%%%%%%%%%%%%%%%%%%%%%%%%%%%%%%%%%%%%%%%%%%%%%%%%%%%%%

To define the supergroup $\SpO(2|1)$ and its action on the supercircle
 we will introduce the notion of functor of points, following \cite{DM}.
 Let $A$ be a supermanifold, an $A$-point of the supercircle is a
 morphism of supermanifolds $A\rightarrow \SII$; we will denote by $\SII(A)$
 the set of $A$-points of $\SII$. The assignation $A\rightarrow
 (A$-points) is the functor of points. An $A$-point of $\SII$ is given by the
 image of the generators $(x,\xi)$ of $\cC^\infty(\SII)$ in
 $\mathcal{O}_A$, the sheaf of functions defining $A$, see \cite{DM, Lei}. By Yoneda's lemma, giving
 $f\in\Diff(\SII)$ is equivalent to giving, functorially in $A$, a map $f_A$
 on $\SII(A)$.
 
 \goodbreak
 
For $\mathcal{A}$ any commutative superalgebra,
 $\GL_{p,q}(\mathcal{A})$ is the well-known group of even invertible linear transformation of
 the free $\mathcal{A}$-module of dimension $p|q$, see \cite{Lei}. We
 define then the supergroup $\GL(p|q)$ by its functor of points,
 $\GL(p|q)(A)=\GL_{p,q}(\mathcal{O}_A)$, and this functor is representable by a
 supermanifold, $\GL(p|q)$. By Yoneda's lemma the action of $\GL(p|q)$
 on $\bbR^{p|q}$ can be given by the action of $\GL(p|q)(A)$ on
 $\bbR^{p|q}(A)$.

If we restrict ourselves to the supermanifolds $A$ whose underlying manifold
 is a point, then $\mathcal{O}_A$ is a Grassmann algebra, and we obtain the
 supermanifolds defined by Rogers \cite{Rog} or the $\mathcal{A}$-manifolds of Tuynman \cite{Tuy}.

From now on we will speak of points instead of $A$-points, and of the action of a supergroup on points, instead of the action of $A$-points of a supergroup on $A$-points.  

The contact structure on $\SII$ (or rather on $\bbR{}P^{1|1}$)
defined by $\a$, see (\ref{contactform}), does stem from the $1$-form on
 $\bbR^{2|1}$ given by $\varpi=\frac{1}{2}(pdq-qdp+\theta d\theta)$, via
  the formula $\varpi=\frac{1}{2}p^2\a$, with $p\neq0$, expressed in affine
 coordinates $x=q/p$ and $\xi=\theta/p$. We define the orthosymplectic
 group \cite{GLS,Man}, denoted by $\SpO(2|1)$, via its functor of points;
 $\SpO(2|1)(A)$ is the
group of all linear transformations of $\bbR^{2\vert1}(A)$, viz.,
\begin{equation}
h=\left(
\begin{array}{ccc} a&b&\g\\ c&d&\d\\ \a&\b&e \\ \end{array} 
\right)
\label{orthosymplectic}
\end{equation}
preserving the symplectic form $d\varpi$, i.e., such that \cite{Man}: 
\begin{eqnarray}
\label{ad-bc=1}
ad-bc-\a\b&=&1,\\
\label{e2=1}
e^2+2\g\d&=&1,\\
\label{alphae}
\a e-a\d+c\g&=&0,\\
\label{betae}
\b e-b\d+d\g&=&0.
\end{eqnarray}
We easily find that $\SpO(2|1)$ also preserves $\varpi$. Since 
$\varpi=\frac{1}{2}p^2\a$, the orthosymplectic group acts by 
contacto\-morphisms, $\SpO(2|1)\rightarrow K(1)$, via the following 
projective action on $\SII$, namely (in terms of $A$-points)
\begin{equation}
\widehat{h}(x,\xi)=\left( \frac{ax+b+\g \xi}{cx+d+\d\xi},\frac{\a 
x+\b+e\xi}{cx+d+\d\xi}\right)
\label{homographie}
\end{equation}
where $h\in\SpO(2|1)$. 

\goodbreak

The Berezinian of $h$ is $\mathrm{Ber}(h)=e+\a\b
 e^{-1}$, see \cite{Man}. We introduce the special orthosymplectic
 group $\SpO_+(2|1)$ as the subgroup of $\SpO(2|1)$ of Berezinian~$1$, or
 as the quotient, $\PC(2|1)$, of $\SpO(2|1)$ by the kernel of the
 projective action (\ref{homographie}), or as the connected component of the
 identity of $\SpO(2|1)$. So, $\SpO_+(2|1)$ is a super-extension of
 $\Sp(2,\bbR)=\SL(2,\bbR)$. We have the following (local) group-factorization
\begin{equation}\label{SpO_decompo}
\SpO_+(2|1)\ni h
=
\left(
\begin{array}{ccc} 1&0&0\\ 
\tilde{c}&1&\tilde{\d}\\ \tilde{\d} &0&1 \\ \end{array} 
\right)
\left(
\begin{array}{ccc} \tilde{a}&0&0\\ 0&\tilde{a}^{-1}&0\\ 0&0&1 \\ 
\end{array} 
\right)
\left(
\begin{array}{ccc} \epsilon&\tilde{b}&-\tilde{\b}\\ 0&\epsilon&0\\ 
0& \epsilon\tilde{\b}&1 \\ \end{array} 
\right)
\label{SpOdecomposition}
\end{equation} 
where 
$(\tilde{a},\tilde{b},\tilde{c},\tilde{\b},\tilde{\d})\in\bbR^{3|2}$,
 with $\epsilon^2=1$, and $\tilde{a}>0$.
Thus, as read off in (\ref{SpO_decompo}), every homography is the composition of an inversion, a dilatation
and a translation. We will denote by $\rE(1|1)$ the subgroup of 
translations and by $\Aff(1|1)$ the subgroup generated by translations
 and dilatations. The connected component of the identity of these subgroups
 of $\SpO_+(2|1)$, characterized by $\epsilon>0$, will be denoted by
 $\rE_+(1|1)$ and $\Aff_+(1|1)$, and referred to as special supergroups. 

\goodbreak

%%%%%%%%%%%%%%%%%%%%%%%%%%%%%%%%%%%%%%%%%%%%%%%%%%%%%%%%%%%%
%%%%%%%%%%%%%%%%%%%%%%%%%%%%%%%%%%%%%%%%%%%%%%%%%%%%%%%%%%%%
\section{Main results}\label{MainResults}
%%%%%%%%%%%%%%%%%%%%%%%%%%%%%%%%%%%%%%%%%%%%%%%%%%%%%%%%%%%%
%%%%%%%%%%%%%%%%%%%%%%%%%%%%%%%%%%%%%%%%%%%%%%%%%%%%%%%%%%%%
We expound in this section the two main results of this paper regarding
 the case of~$\SII$; the first one gives the invariants of the action on $\SII$
 of the special supergroups $\rE_+(1|1)$, $\Aff_+(1|1)$ and $\SpO_+(2|1)$, and the second one provides,  by means of a super version of the Cartan formula, the associated $K(1)$-cocycles.
 These results will be extended (whenever possible) to the case of $\rS^{1\vert N}$ in Section \ref{S1N}. 

%%%%%%%%%%%%%%%%%%%%%%%%%%%%%%%%%%%%%%%%%%%%%%%%%%%%%%%%%%%%
\subsection{Super Euclidean, affine and projective invariants} 
%%%%%%%%%%%%%%%%%%%%%%%%%%%%%%%%%%%%%%%%%%%%%%%%%%%%%%%%%%%%

Let $t_1,t_2,t_3,t_4$ be four generic points of $\SII$, 
$t_i=(x_i,\xi_i)$. 
\begin{thm}\label{thminvariant}
The following three couples, $I_\mathsf{e}$, $I_\mathsf{a}$ and
 $I_\mathsf{p}$, of superfunctions are the invariants of the action of
 Euclidean, affine and projective special supergroups on $\SII$:
\begin{itemize}
\item Euclidean invariant:
 $I_\mathsf{e}(t_1,t_2)=([t_1,t_2],\{t_1,t_2\})$ with
\begin{eqnarray}\label{Einv}
[t_1,t_2]&=&x_2-x_1-\xi_2\xi_1, \\[6pt]
 \{t_1,t_2\}&=&\xi_2-\xi_1.
\end{eqnarray}
\item Affine invariant, $I_\mathsf{a}(t_1,t_2,t_3)=([t_1,t_2,t_3],\{t_1,t_2,t_3\})$, where, if $x_1<x_2$,
\begin{eqnarray}\label{Ainv}
[t_1,t_2,t_3]
&=&
\frac{[t_1,t_3]}{[t_1,t_2]}, \\[6pt]
\{t_1,t_2,t_3\}
&=&
[t_1,t_2,t_3]^{\half}\,\frac{\{t_1,t_3\}}{[t_1,t_3]^\half}.
\end{eqnarray}
\item Projective invariant, $I_\mathsf{p}(t_1,t_2,t_3,t_4)=([t_1,t_2,t_3,t_4],\pm\{t_1,t_2,t_3,t_4\})$, where, when $\ord(t_1,t_2,t_3)=1$, see (\ref{ord}),
\begin{eqnarray}
\label{OurSuperCrossRatio}
[t_1,t_2,t_3,t_4]&=&\frac{[t_1,t_3][t_2,t_4]}{[t_2,t_3][t_1,t_4]},
 \\[6pt] 
\label{projectiveinvariant}
\{t_1,t_2,t_3,t_4\}&=&[t_1,t_2,t_3,t_4]^\half\;
 \frac{\{t_2,t_4\}[t_1,t_2]-\{t_1,t_2\}[t_2,t_4]}{([t_1,t_2][t_2,t_4][t_1,t_4])^\half}.
\end{eqnarray} 
\end{itemize} 
If a bijective transformation of $\SII$ preserves one of these three couples of
super\-functions, it can be identified with the action of an element of 
the corresponding supergroup, $\rE_+(1|1)$, 
$\Aff_+(1|1)$ or $\SpO_+(2|1)$. Moreover, if a contactomorphism
 $\Phi\in{}K(1)$ preserves the even part of one of the invariants
 $I_\mathsf{e}$, $I_\mathsf{a}$, or $I_\mathsf{p}$, respectively, then $\Phi=\widehat{h}$ for some
 $h$ in $\rE(1|1)$, $\Aff(1|1)$, or $\SpO_+(2|1)$, respectively.
\end{thm} 
This theorem summarizes Theorems \ref{EuclideanThm}, \ref{AffineThm},
 and \ref{ProjectiveThm} given below, as well as their corollaries.
Their proofs  rely on the $p|q$-transitivity of the action of these supergroups on
 $\SII$; all details are given in Section \ref{InvariantSection}.

\begin{rmk}
{\rm
The super cross-ratio, i.e., the even part (\ref{OurSuperCrossRatio}) of the projective invariant, $I_\mathsf{p}$, has already been introduced by Nelson \cite{Nel}, and
 used by Giddings \cite{Gid} while studying the punctured super Riemann
 sphere, and also by Uehara and Yasui \cite{UY} to define coordinates on
 the super Teichm\"uller space. It has also been put forward by Manin in
 \cite{Man} from a somewhat different standpoint that we can summarize
 as follows in our formalism. Using the even 
symplectic form $d\varpi=dp\wedge dq+\frac{1}{2}d\theta\wedge d\theta$
 on~$\bbR^{2|1}$ one 
defines a $\SpO_+(2|1)$-invariant pairing  
$\left\langle Z_i,Z_j\right\rangle = d\varpi
 (Z_i,Z_j)=p_ip_j[t_j,t_i]$, for $Z_i=(p_i\; q_i 
\;\theta_i)\in\bbR^{2|1}$, where $t_i=(q_i/p_i,\theta_i/p_i)$. Positing
 $[Z_1,Z_2,Z_3,Z_4]=\frac{\left\langle 
Z_3,Z_1\right\rangle\left\langle Z_4,Z_2\right\rangle}{\left\langle 
Z_3,Z_2\right\rangle\left\langle Z_4,Z_1\right\rangle}$, one obtains a
 four-point function, not only 
$\SpO_+(2|1)$-invariant, but also invariant under rescalings of each
 variable. We then have $[Z_1,Z_2,Z_3,Z_4]=[t_1,t_2,t_3,t_4]$, see~(\ref{OurSuperCrossRatio}).
}
\end{rmk}

\begin{rmk}
{\rm
The odd part (\ref{projectiveinvariant}) of the projective invariant,
 $I_\mathsf{p}$, can clearly be reduced to a three-point (almost)
 invariant function, corresponding to $J_\mathsf{p}$ given below in (\ref{epsilon}).
 The latter was already introduced by D'Hoker and Phong \cite{DHP} and
 used in \cite{Gid, UY} on the same footing as the super cross-ratio. We
 have written $J_\mathsf{p}$ as function of the Euclidean invariants, but it can be
 recast into the form
\begin{equation}\label{odd_Ip_cyclic}
J_\mathsf{p}(t_1,t_2,t_3)=\pm\frac{\xi_1[t_2,t_3]+\xi_2[t_3,t_1]+\xi_3[t_1,t_2]-\xi_1\xi_2\xi_3}{([t_1,t_3][t_3,t_2][t_2,t_1])^\half},
\end{equation}
which precisely corresponds to the expression 
originally given in \cite{Aok,DHP}, where the cyclic symmetry is
 obvious. This invariant, $J_\mathsf{p}$, has also been introduced by Manin in
 \cite{Man}, using a construction akin to that developed by us in Section \ref{InvariantSection}.
}
\end{rmk}

\begin{rmk}
{\rm
If we apply the projection $\pi:\Cinfty(\SII)\to\Cinfty(S^1)$, see
 (\ref{quotient}), to each invariant $I_\mathsf{e}$, $I_\mathsf{a}$ and $I_\mathsf{p}$, we obtain the usual Euclidean, affine and projective invariant, namely the distance, the distance-ratio and the cross-ratio.  
}
\end{rmk}
 
%%%%%%%%%%%%%%%%%%%%%%%%%%%%%%%%%%%%%%%%%%%%%%%%%%%%%%%%%%%%
\subsection{The associated $1$-cocycles of $K(1)$}
%%%%%%%%%%%%%%%%%%%%%%%%%%%%%%%%%%%%%%%%%%%%%%%%%%%%%%%%%%%%

Let $\Phi\in \Diff(\rS^1)$ be a diffeomorphism of the circle, and
 $\phi_{\ve}=\Id+\ve X+O(\ve^2)$ 
be the flow of a vector field $X$ on the circle. We set
 $t_i=\phi_{(i-1)\ve}(t_1)$ for $i=1,2,3,4$. 
Then, the Schwarzian derivative can be defined in terms of the 
cross-ratio, as the quadratic differential $\cS(\Phi)\in\cQ(\rS^1)$
 appearing in the Cartan formula, see \cite{Car,OT}: 
\begin{equation}
\frac{\Phi^*[t_1,t_2,t_3,t_4]}{[t_1,t_2,t_3,t_4]}-1=\left\langle \ve 
X\otimes \ve X, \cS(\Phi)\right\rangle + O(\ve^3).
\label{CartanFormula}
\end{equation}

For the group of contactomorphisms of $\SII$, we will proceed by
 analogy with this method. Starting from the super cross-ratio
 (\ref{OurSuperCrossRatio}), we will deduce the super Schwarzian derivative,
 $\cS(\Phi)\in\cQ (\SII)$, as a $K(1)$-cocycle with kernel 
$\SpO_+(2|1)$. Euclidean and affine $K(1)$-cocycles will, likewise, be
 obtained from the even Euclidean and affine invariants. We recall that
 $\Omega^1(\SII)$ is the space of 
$1$-forms, $\cQ(\SII)$ the space of quadratic differentials of the
 supercircle, and 
$E_\Phi=\frac{\Phi^*\a}{\a}=(D\psi)^2$, see Subsection~\ref{ContactS11}.

\goodbreak

\begin{thm}\label{thmcocycle}
From the Euclidean (\ref{Einv}), affine (\ref{Ainv}), and projective
 (\ref{OurSuperCrossRatio}) even invariants, we deduce via the Cartan
 formula (\ref{CartanFormula}) three $1$-cocycles of $K(1)$, with kernel 
$\rE(1|1)$, $\Aff(1|1)$ and $\SpO_+(2|1)$ respectively. They retain the
 following form:
\begin{itemize}

\item the Euclidean cocycle $\cE:K(1)\rightarrow \cF_0(\SII)$:
\begin{equation}
\label{Euclideancocycle}
\cE(\Phi)=\log E_\Phi=\log(D\psi)^2,
\end{equation}
\item 
the affine cocycle $\cA:K(1)\rightarrow \Omega^1(\SII)$:
\begin{equation}
\label{Affinecocycle}
\cA(\Phi)=d\cE(\Phi),%=\frac{dE_\Phi}{E_\Phi}
\end{equation}

\item the projective Schwarzian cocycle $\cS:K(1)\rightarrow
 \cQ(\SII)$:
\begin{equation}
\label{Projectivecocycle}
\cS(\Phi)
%%% =
%%% \frac{1}{6}\alpha^2\; D\widetilde{\rS}(\Phi)
%%% +\frac{1}{2}\alpha\beta\; \widetilde{\rS}(\Phi)\\
=\frac{2}{3}\,\alpha^\half L_D\,\rS(\Phi),
\end{equation}
\end{itemize}
where $L_D$ stands for the Lie derivative with respect to the vector
 field $D$, and $\rS(\Phi)$ is given by
%%% $\widetilde{\rS}(\Phi)=\rS(\Phi)\a^{-3/2}$, see 
Equation (\ref{S_1}) below.
Moreover, using the projections on tensor densities defined in
 Proposition \ref{split}, we obtain two new affine and projective $1$-cocycles,
 namely
\begin{itemize}
%%% \item The projection of the Euclidean cocycle is trivial.
\item the projection of the affine cocycle, $\rA:K(1)\rightarrow
 \cF_\half(\SII)$:
\begin{equation}
\label{DAffinecocycle}
\rA(\Phi)=\a^\half\left\langle
 D,\cA(\Phi)\right\rangle=\frac{DE_\Phi}{E_\Phi}\;\a^\half,
\end{equation}

\item the projection of the Schwarzian cocycle, $\rS:K(1)\rightarrow
 \cF_{3/2}(\SII)$:
\begin{equation}
\label{DProjectivecocycle}
\rS(\Phi)=\a^\half\left\langle D,\cS(\Phi)
 \right\rangle=\frac{1}{4}\left(\frac{D^3E_\Phi}{E_\Phi}-\frac{3}{2}\frac{DE_\Phi\,D^2E_\Phi}{E_\Phi^2}
 \right)\alpha^{3/2}.
\label{S_1}
\end{equation}
\end{itemize}
\end{thm}
We will give the proof of this theorem in Section \ref{proofcocycle}.

\begin{rmk}
{\rm
As in the case of the Schwarzian cocycle
 (\ref{Projectivecocycle}), using Proposition~\ref{split}, we can express the affine cocycle $\cA$ in terms of its projection
 $\rA$, namely 
\begin{equation}
\cA(\Phi)=\a^\half L_D \, \rA(\Phi). 
\label{newcA}
\end{equation}
}
\end{rmk}

\begin{rmk}\label{rmkprojection}
{\rm
1) The projection $\pi:\Cinfty(\SII)\to\Cinfty(S^1)$, see
 (\ref{quotient}), can be extended naturally to differential forms and quadratic
 differentials, sending $\a$ to $dx$ and $\b$ to~$0$. So, we can project the
 $K(1)$-cocycle $\cS(\Phi)$ given by (\ref{Projectivecocycle}) on
 $\cQ(S^1)$, and as the result depends only on $f=\Pi(\Phi)$, see
 (\ref{diagram}), we easily recover the classical Schwarzian derivative $\rS_0:\Diff_+(\rS^1)\mapsto \cQ(\rS^1)$, namely
\begin{equation}
\label{classicSch}
\rS_0(f)=\left(\frac{f'''}{f'}-\frac{3}{2}\left(\frac{f''}{f'}\right)^2\right)dx^2,
\end{equation}
using the expression (\ref{devt-cr}) where $\pi(E_\Phi)=f'$. See, e.g., \cite{Car,DG,OT}. The projections of the two other $K(1)$-cocycles, $\cE$ and $\cA$, lead to the Euclidean and affine cocycle of $\Diff_+(\rS^1)$, namely $\cE_0(f)=f'$ and $\cA_0(f)=\frac{f''}{f'}dx$.

2) The $K(1)$-cocycle, $\rS$, given in (\ref{S_1}), is the super
 Schwarzian derivative, independently introduced by Friedan \cite{Fri} and
 Radul  \cite{Rad}. Recall that $E_\Phi=(D\psi)^2$, see~(\ref{E}), so we
 can also write
\begin{equation}
\rS(\Phi)=\left(\frac{D^4\psi}{D\psi}-2\frac{D^2\psi\,D^3\psi}{(D\psi)^2} \right)\alpha^{3/2}.
\end{equation} 
This is the form of the super Schwarzian derivative used in
 superconformal field theories~\cite{Fri}, see also \cite{Man}. Gieres and Theisen
 use it in \cite{Gie}, as well as the affine cocycle $A$, to construct
 superconformal covariant operators.
}
\end{rmk}

It is well-known that the classical Schwarzian derivative
 (\ref{classicSch}) can be expressed in terms of the classical affine cocycle
 $\rA_0(f)=(f''/f')dx$ on $S^1$, viz.,
\begin{equation}\label{A-S}
\rS_0(f)
=
dx\, L_{\partial_x}\rA_0(f) -\half\rA_0(f)^2,
\end{equation} 
where $f\in \Diff_+(\rS^1)$. A formula relating, in the super case, the
 expression of $\rS$ and $\rA$ can be found in \cite{Gie}. The next
 proposition gives another formula for the $1$-cocycle $\rS$ in a form akin
 to (\ref{A-S}).

\begin{pro}\label{cArS}
Let $\cA$ denote the affine $K(1)$-cocycle (\ref{Affinecocycle}); the
 following holds true for the super Schwarzian derivative (\ref{S_1}):
\begin{equation}
\rS(\Phi)=\frac{1}{4}\,\a^\half \Big\la D, (\a^\half L_D)^2 \cA(\Phi)
 -\half \cA(\Phi)^2\Big\ra.
\label{NiceFormula}
\end{equation} 
\end{pro}

%%%%%%%%%%%%%%%%%%%%%%%%%%%%%%%%%%%%%%%%%%%%%%%%%%%%%%%%%%%%
\subsection{The determination of $H^1(K(1),\cF_\l)$}
%%%%%%%%%%%%%%%%%%%%%%%%%%%%%%%%%%%%%%%%%%%%%%%%%%%%%%%%%%%%

The following corollary of Theorem \ref{thmcocycle} is straightforward; its proof relies on the expres\-sion (\ref{e(Xf)}) of the Euclidean $1$-cocycle of $k(1)$, the Lie superalgebra of infinitesimal contactomorphisms of $\SII$.
\begin{cor}
The Lie algebra $1$-cocycles associated with the $K(1)$-cocycles $\cE$,
 $\rA$, and~$\rS$, read $c_i:k(1)\rightarrow\cF_{i/2}(\SII)$, with
\begin{equation}
\label{Algebrecocycle}
c_i(X_f)= (D^{i+2} f)\; \a^{i/2},
\end{equation}
where $i=0,1,3$. 
\end{cor}
We recover, in this way, three of the four nontrivial $1$-cocycles of $k(1)$
 with coefficients in $\cF_\l$ (see \cite{ABF} for a classification).
 The fourth one, $\tilde{c}_0:k(1)\rightarrow \cF_0(\SII)$, defined by 
$\tilde{c}_0(X_f)=f-\frac{1}{2}\xi\partial_\xi f$, does not integrate
 as a group $1$-cocycle, just like the $\Vect(\rS^1)$-cocycle $X_f \mapsto f$. Indeed, suppose that $\tilde{c}_0$ does integrate as a $K(1)$-cocycle, $\tilde{C}_0$. Then $\partial_x\in k(1)$ induces, using an angular coordinate $x$, the
 flow $\Phi_t(x,\xi)=(x+t,\xi)$, and as $\tilde{c}_0(\partial_x)=1$, we
 have $\tilde{C}_0(\Phi_t)=\half t$, see e.g. \cite{Sou}. But this is inconsistent with the periodicity condition $\Phi_t=\Phi_{t+2\pi}$. This is a straightforward generalization to the super-algebraic framework of the observation \cite{OR} that the only $\Vect(S^1)$ $1$-cocycles that integrate as $\Diff_+(S^1)$-cocycles are those which are Euclidean-basic; see also \cite{GR}. Here, one checks that $\tilde{c}_0$ is not $\rE(1|1)$-basic. 
 
 As the derivation of Lie group cocycle is an
 injection from the Lie group cocycle into the Lie algebra cocycle, we obtain
 the complete classification of the nontrivial $1$-cocycles of $K(1)$ with values in $\cF_\l$.
 
\begin{thm}\label{ThmH1}
\begin{enumerate}
\item
The cohomology spaces $H^1(K(1),\cF_\l)$ are given by
\begin{equation}
H^1(K(1),\cF_\l)=
\left\{
\begin{array}{ll}
\bbR &\mbox{  if\ }\l=0,\,\frac{1}{2},\,\frac{3}{2}\\[6pt]
\{0\} &\mbox{ otherwise.}
\end{array}
\right.
\label{H1Fl}
\end{equation}
These three cohomology spaces are respectively generated by $\cE$, 
$\rA$ and $\rS$. 

\item
Moreover, the two cohomology spaces
\begin{eqnarray}
\label{H1Omega}
H^1(K(1),\Omega^1(\SII))&=&\bbR, \\[6pt]
\label{H1Q}
H^1(K(1),\cQ(\SII))&=&\bbR,
\end{eqnarray} 
are respectively generated by $\cA$ and $\cS$.
\end{enumerate}
\end{thm}
\begin{proof}
We have already proved (\ref{H1Fl}) in the course of the above discussion. Let us now derive (\ref{H1Omega}). Suffice it to notice that Proposition \ref{split} yields the decomposition $\Omega^1(\SII)=\cF_\half\oplus\cF_1$ into $K(1)$-submodules. The classification~(\ref{H1Fl}) then shows that the image by the section $\alpha^\half{}L_D$ of the generator $A$ of $H^1(K(1),\cF_\half)$ spans $H^1(K(1),\Omega^1(\SII))$. The same argument holds for the proof of (\ref{H1Q}).
\end{proof}
Let us end up with the following synthesis of the results obtained in this section.

\goodbreak

\begin{rmk}\label{rmkN=1}
{\rm
We have thus established a $1$-$1$ correspondence between the set of nontrivial cohomology spaces $H^1(K(1),\cF_\l)$ (or $H^1(K(1),\cM)$, with $\cM=\Omega^0(\SII)$, $\Omega^1(\SII)$, $\cQ(\SII)$) and the ``natural'' geometries of the supercircle, namely the Euclidean, affine, and projective geometries of $\SII$. These geometries are defined by the kernels of the corresponding $1$-cocycles $\cE,\cA,\cS$. These groups, in turn, give rise to the invariants $I_\mathsf{e},I_\mathsf{a},I_\mathsf{p}$. At last, these invariants lead us back to the generators of the above cohomology spaces, with the help of the Cartan-like formul{\ae} (\ref{CartanEuclid}), (\ref{CartanAffine}), and (\ref{CartanFormula}). 
}
\end{rmk}

%%%%%%%%%%%%%%%%%%%%%%%%%%%%%%%%%%%%%%%%%%%%%%%%%%%%%%%%%%%%
%%%%%%%%%%%%%%%%%%%%%%%%%%%%%%%%%%%%%%%%%%%%%%%%%%%%%%%%%%%%
\section{Super Euclidean, affine and projective invariants of $\SII$}
\label{InvariantSection}
%%%%%%%%%%%%%%%%%%%%%%%%%%%%%%%%%%%%%%%%%%%%%%%%%%%%%%%%%%%%
%%%%%%%%%%%%%%%%%%%%%%%%%%%%%%%%%%%%%%%%%%%%%%%%%%%%%%%%%%%%
In this section we construct the Euclidean, affine and projective 
invariants given by Theorem \ref{thminvariant}. We introduce an
 extension of the notion of transitivity, allowing us to formulate a theorem
 giving the sought invariants when applied to each supergroup: $\rE_+(1|1)$,
 $\Aff_+(1|1)$, and $\SpO_+(2|1)$.

Let us first introduce an equivalence relation, on the $n$-tuples of a
 product set $E=E_0\times E_1$. We denote by $p_0$ and $p_1$ the two
 canonical projections. Let $s=(s_1,\ldots,s_n)$ and $t=(t_1,\ldots,t_n)$ be two
 $n$-tuples of $E$, we will say that $s$ and $t$ are $p|q$ equivalent, $s\stackrel{p|q}{=}t$, where $n=\max(p,q)$, iff
\begin{equation}\label{p|q-transitive}
\forall i\in\llbracket 1,p\rrbracket, \quad p_0(s_i)=p_0(t_i) \qquad
 \text{and}\qquad \forall i\in\llbracket 1,q\rrbracket, \quad
 p_1(s_i)=p_1(t_i).
\end{equation} 
We will use the notation $[t]$ for the class of $t$ for this equivalence relation.

\goodbreak

\begin{defi}
Let $G$ be a group acting on a set $E=E_0\times E_1$ by ($g\mapsto
 \widehat{g}$). The action of $G$ on $E$ is $p|q$-transitive, $n=\max(p,q)$,
 if for any $n$-tuples $s$ and $t$ of distinct points, there exists an element $h\in G$
 such that $\widehat{h}(t)\stackrel{p|q}{=}s$. If $h$ is unique the action is said to be simply $p|q$-transitive.
\end{defi}
In particular a $p|q$-transitive action is $\min(p,q)$-transitive.
To prove $n$-transitivity, we usually prove that any $n$-tuple $t$ can
 be sent to a given $n$-tuple $m$. To prove $p|q$-transitivity we need an extra condition, this is specified by the next
 proposition.

\begin{pro}\label{pro_pq-transitive}
Let $G$ act on a set $E=E_0\times E_1$ and choose $m$, a $n$-tuple of~$E$. Suppose that for every $n$-tuple $s$, there exists
 $h\in G$ such that $\widehat{h}(s)\stackrel{p|q}{=}m$, where $n=\max (p,q)$, and
 $G.[s]\supseteq [m]$. Then the action of $G$ on $E$ is $p|q$-transitive.      
\end{pro}

\begin{proof}
Let $t$ and $s$ be two $n$-tuples of $E$. We look for those $k\in G$ such
 that $\widehat{k}(t)\stackrel{p|q}{=}s$. By assumption, there exist
 $h,g\in G$ such that $\widehat{h}(t)\stackrel{p|q}{=}m$ and
 $\widehat{g}(s)\stackrel{p|q}{=}m$. Then, as $\widehat{h}(t)\in [m]$ and
 $G.[s]\supseteq [m]$, there exist $s'\stackrel{p|q}{=}s$ and $g'\in G$ such that
 $\widehat{g'}(s')=\widehat{h}(t)$. Finally $\widehat{g'}^{-1}(\widehat{h}(t))\stackrel{p|q}{=}s$.   
\end{proof}

\begin{thm}
Let $g\mapsto \widehat{g}$ denote the simply $p|q$-transitive action of
 a group $G$ on a set $E=E_0\times E_1$, 
and let $m$ be a $n$-tuple, $n=\max(p,q)$, of distinct points of $E$. We
 can define the following $(n+1)$-point function of $E$ with values in $E$, associated to the class of $m$, namely
\begin{equation}\label{I_m}
I_{[m]}(t_1,\ldots,t_{n+1})=\widehat{h}(t_{n+1})
\end{equation}
where $\widehat{h}(t)\stackrel{p|q}{=}m$, and $t=(t_1,\ldots,t_n)$ is a $n$-tuple of distinct points of $E$.
This function enjoys the following properties:
\begin{enumerate}
\item 
$I_{[m]}$ is $G$-invariant.
\item 
If $\Phi \in E!$ preserves $I_{[m]}$, then $\Phi=\widehat{g}$ for some~$g\in 
G$.
\item 
Let $l$ be a $n$-tuple of $E$ and $g\in G$, then $\widehat{g}[m]=[l]$
 iff $I_{[l]}=\widehat{g}\circ I_{[m]}$.    
\end{enumerate}
The first two properties assert that $I_{[m]}$ is a characteristic
 invariant of the action of $G$.\\
Moreover, if $n=p>q$, we can define $n$-point invariant functions with
 values in $E_1$ 
\begin{equation}\label{Jmj}
J_{[m],j}(t)=p_1(\widehat{h}(t_j))
\end{equation}
for $j\in\llbracket q+1,p \rrbracket$. Any $(n+1)$-point $G$-invariant
 function $I$ can be factorized through the invariants $I_{[m]}$ and
 $J_{[m],j}$, i.e., $I=f(J_{[m],q+1},\ldots,J_{[m],p},I_{[m]})$ for some function~$f$, depending on the $n$-tuple $m$. Similarly, any $n$-point $G$-invariant
 function can be factorized through the invariants $J_{[m],j}$.
\label{generalinvariant}
\end{thm}

\begin{proof}
We first prove that $I_{[m]}(\widehat{g}(t_1),\ldots,\widehat{g}(t_{n+1}))=I_{[m]}(t_1,\ldots,t_{n+1})$ for all $g\in G$, i.e., $I_{[m]}$ is $G$-invariant. 
Since $\widehat{h}(t)\stackrel{p|q}{=}m$, we have $\widehat{h}\circ 
\widehat{g}^{-1}(\widehat{g}(t))\stackrel{p|q}{=}m$. It follows that 
$I_{[m]}(\widehat{g}(t_1),\ldots,\widehat{g}(t_{n+1}))=
 \widehat{h}\circ 
\widehat{g}^{-1}(\widehat{g}(t_{n+1}))=\widehat{h}(t_{n+1})$, hence the result. The proof of the $G$-invariance of $J_{[m],j}$ is identical.

Secondly, we show that $I_{[m]}$ is a characteristic 
$G$-invariant. Let $\Phi$ be a bijection of $E$, such that $\Phi^*I_{[m]}=I_{[m]}$,
 we have to prove that 
$\Phi$ comes from an element of $G$. There exist $h,g\in G$, depending
 on $t$ such that, $\widehat{h}(t)\stackrel{p|q}{=}m$ and
 $\widehat{g}(\Phi(t))\stackrel{p|q}{=}m$. Since $\Phi^*I_{[m]}=I_{[m]}$, we have
 $\widehat{g}(\Phi(t_{n+1}))=\widehat{h}(t_{n+1})$ for all $t_{n+1}\in E$, and thus 
$\Phi=\widehat{k}$ , with $k=g^{-1} h$.

Thirdly, suppose that there exists $g\in G$ such that
 $\widehat{g}[m]=[l]$. Let $t$ be a $n$-tuple, we have
 $\widehat{h}(t)\stackrel{p|q}{=}m$ for a unique $h\in G$, then
 $\widehat{g}(\widehat{h}(t))\stackrel{p|q}{=}l$, and it follows that 
$I_{[l]}(t_1,\ldots,t_{n+1})=\widehat{g}\circ\widehat{h}(t_{n+1})$.
Conversely suppose that $I_{[l]}=\widehat{g}\circ I_{[m]}$ for some
 $g\in G$ and let $m'\in[m]$. For every $n$-tuple $t$, there exists $h\in
 G$ such that $\widehat{h}(t)\stackrel{p|q}{=}m$ and then $I_{[l]}(t,t_{n+1})=\widehat{g}\circ\widehat{h}(t_{n+1})$, for all $t_{n+1}\in E$. As $I_{[l]}(t,t_{n+1})=\widehat{k}(t_{n+1})$ for the unique $k\in G$ such that $\widehat{k}(t)\stackrel{p|q}{=}l$, we deduce that $\widehat{g}(\widehat{h}(t))\stackrel{p|q}{=}l$. In particular for the $n$-tuple $m'$, $h$
 is the identity, hence $\widehat{g}(m')\stackrel{p|q}{=}l$. It
 follows that $\widehat{g}[m]\subseteq [l]$, and as we also have
 $\widehat{g^{-1}}\circ I_{[l]}=I_{[m]}$, then $\widehat{g^{-1}}[l]\subseteq [m]$,
 leading to the result $\widehat{g}[m]=[l]$.

Fourthly, let $I$ be an arbitrary $(n+1)$-point invariant function.
 For any $n$-tuple~$t$ there exists some $h\in G$ such that
 $I(t_1,\ldots,t_{n+1})=I(m'_1,\ldots,m'_n,\widehat{h}(t_{n+1}))$ with
 $\widehat{h}(t)=m'\stackrel{p|q}{=}m$. Now
 $I_{[m]}(t_1,\ldots,t_{n+1})=\widehat{h}(t_{n+1})$ and since $m'$ depends only on $m$ and on $J_{[m],j}$, the result follows. 
\end{proof}   

\begin{rmk}\label{transitive}
{\rm
This theorem generalizes the more common situation of a simply
 $n$-transitive action of a group, choosing $p=q$. In this case, $\stackrel{p|q}{=}$ reduces to the mere equality, $=$, and every
 $(n+1)$-point invariant can be factorized through the invariant~$I_m$ given by
 Theorem \ref{generalinvariant}. In particular, the invariant $I_l$, for
 $l$ another $n$-tuple, can be factorized $I_l=\widehat{g}\circ I_m$, with
 $g\in G$  such that $\widehat{g}(m)=l$.
}
\end{rmk}
\begin{rmk}
{\rm
In the definition of $p|q$-transitivity and in this theorem, we consider $n$-tuples of distinct points. The notion of distinct points of $E=E_0\times E_1$ is well-known, but we will strengthen it by assuming distinct even coordinates when dealing with supergroups acting on the supercircle.
}
\end{rmk}
As direct and classical application of our result, the action of $\PGL(2,\bbR)$ by homographies on the circle $\rS^1$, viewed as $\bbR P^1$, is simply $3$-transitive, and choosing
 $m=(\infty,0,1)$ as the distinguished triple of points, we obtain the usual cross-ratio:
 $I(x_1,x_2,x_3,x_4)=\frac{(x_1-x_3)(x_2-x_4)}{(x_2-x_3)(x_1-x_4)}$.
 
 \goodbreak

%%%%%%%%%%%%%%%%%%%%%%%%%%%%%%%%%%%%%%%%%%%%%%%%%%%%%%%%%%%%
\subsection{Euclidean invariants}
%%%%%%%%%%%%%%%%%%%%%%%%%%%%%%%%%%%%%%%%%%%%%%%%%%%%%%%%%%%%

We introduce the subgroups $\rE(1|1)$ and $\rE_+(1|1)$ of $\SpO_+(2|1)$
 which act on 
$\SII$ by translations in an affine coordinate system.
\begin{defi}\label{defE11}
Let us define $\rE(1|1)$ as the subgroup  of $\GL(2|1)$ whose elements
 are of the form 
\begin{equation}
g=\left(
\begin{array}{ccc} 
\epsilon&\epsilon b&-\epsilon\b\\ 
0&\epsilon&0\\ 
0&\b&1 \\ 
\end{array} 
\right)
\label{E11}
\end{equation}
where $(b,\b)\in\bbR^{1|1}$, and $\epsilon^2=1$. It acts on $\bbR^{1|1}\subset\SII$ by translations, according to $\widehat{g}(x,\xi)=(x+b-\b\xi, \epsilon\b+\epsilon\xi)$.
  We will denote by $\rE_+(1|1)$ the connected component of the
 identity characterized by $\epsilon=1$.
\end{defi} 

\begin{rmk}
{\rm
The Euclidean groups can be defined in an alternative manner, in terms of the transformation laws of the $1$-forms $\a$ and $\b$, and then directly as subgroups of $K(1)$. The group $\rE(1|1)$ is the subgroup of those $\Phi\in\Diff(\SII)$ such that $\Phi^*\a=\a$ and $\Phi^*\b=\epsilon\b$, with $\epsilon=\pm 1$; restricting to $\epsilon =1$ we obtain the subgroup $\rE_+(1|1)$. 
}
\end{rmk}

\begin{pro}\label{EuclideanThm}
The action of $\rE_+(1|1)$ on $\bbR^{1|1}\subset\SII$ is simply $1|1$-transitive;
 choosing $\mathsf{e}=(0,0)$, it defines a characteristic Euclidean invariant
 consisting of the following two-point couple of superfunctions
\begin{equation}\label{I_euclidien}
I_\mathsf{e}(t_1,t_2)=([t_1,t_2]\,
 ,\,\{t_1,t_2\})=(x_2-x_1-\xi_2\xi_1\, 
,\,\xi_2-\xi_1)
\label{euclideaninvariant}
\end{equation} 
where $t_1=(x_1,\xi_1)$ and $t_2=(x_2,\xi_2)$.
\end{pro}
\begin{proof}
Following Theorem \ref{generalinvariant}, we have to show that for any 
point $t_1$ of $\SII$, there exists a unique $h\in\rE_+(1|1)$ such 
that $\widehat{h}(t_1)=(0,0)$, and then to compute 
$\widehat{h}(t_2)=([t_1,t_2]\, ,\,\{t_1,t_2\})$ for another point
 $t_2$. 

The action of any $h\in\rE_+(1|1)$ is given by 
$\widehat{h}(x,\xi)=(x+b-\b\xi, \b+\xi)$. Hence
 $\widehat{h}(t_1)=(0,0)$ is equivalent to 
$x_1+b-\b\xi_1=0$ and $\b+\xi_1=0$, i.e., $\b=-\xi_1$ and $b=-x_1$. So, $h$ is 
uniquely determined, and
 $\widehat{h}(t_2)=(x_2-x_1-\xi_2\xi_1,\xi_2-\xi_1)$, 
as announced.
\end{proof}
The choice of the point $e=(0,0)$ is immaterial, see Remark
 \ref{transitive}. 

\begin{rmk}
{\rm
The even Euclidean invariant $[t_1,t_2]$ is the discretized version of
 the 
contact form $\a=dx+\xi d\xi$, while the odd Euclidean invariant 
$\{t_1,t_2\}$ is that of $\b=d\xi$. This will be specified in Lemma
 \ref{lemdiscret}.
}
\end{rmk}

\begin{cor}\label{eveneucl}
The even part of $I_\mathsf{e}$ is invariant under $\rE(1|1)$, and
 characterizes this subgroup of $K(1)$, namely if $\Phi\in K(1)$ satisfies
 $\Phi^*[t_1,t_2]=[t_1,t_2]$, then $\Phi=\widehat{h}$ for some $h\in\rE(1|1)$.
\end{cor}
\begin{proof}
Let $\iota\in K(1)$ be defined by $\iota:(x,\xi)\mapsto (x,-\xi)$.
 Identifying $\rE(1|1)$ with its image in $K(1)$ we have
 $\rE(1|1)=\rE_+(1|1)\sqcup \iota(\rE_+(1|1))$. Since $[t_1,t_2]$ is invariant under $\rE_+(1|1)$
 as well as under the action of $\iota$, this is a $\rE(1|1)$-invariant. 

Let $\Phi=(\varphi,\psi)\in K(1)$ be such that
 $\Phi^*[t_1,t_2]=[t_1,t_2]$. There exists $t_1$ such that $\Phi(t_1)=(0,0)$, and
 $h\in\rE_+(1|1)$ such that $\widehat{h}(t_1)=\Phi(t_1)$. Since $\Phi$ leaves
 $[t_1,t_2]$ invariant, we have
 $\varphi(t_2)=[\Phi(t_1),\Phi(t_2)]=[t_1,t_2]=\widehat{h}_0(t_2)$ in view of (\ref{I_m}); hence $\varphi=\widehat{h}_0$, with
 $\widehat{h}=(\widehat{h}_0,\widehat{h}_1)$. Using Lemma \ref{even-all}, we
 obtain $\Phi=\widehat{h}$ or $\Phi=\iota(\widehat{h})$, and $\Phi$ is then
 a (super) translation.
\end{proof}
%%%%%%%%%%%%%%%%%%%%%%%%%%%%%%%%%%%%%%%%%%%%%%%%%%%%%%%%%%%%
\subsection{Affine invariants} 
%%%%%%%%%%%%%%%%%%%%%%%%%%%%%%%%%%%%%%%%%%%%%%%%%%%%%%%%%%%%

Let us start with the definitions of $\Aff(1|1)$ and $\Aff_+(1|1)$ and with their action on $\SII$.

\goodbreak

\begin{defi}
The affine supergroup, $\Aff(1|1)$, is the subgroup  of $\GL(2|1)$ whose
 elements are of the form 
\begin{equation}
g=\left(\begin{array}{ccc} a&ab&-a\b\\ 0&a^{-1}&0\\ 
0&\b&1 \\ \end{array} \right)
\label{A11}
\end{equation}
where $(a,b,\b)\in\bbR^{2|1}$, and $a\neq0$. This supergroup acts on $\bbR^{1|1}\subset\SII$
 by translations and 
dilatations, $\widehat{g}(x,\xi)=(a^2x+a^2 b-a^2\b\xi, a\b+a\xi)$.  
We will denote by $\Aff_+(1|1)$ the connected component of the identity, characterized by $a>0$.
\end{defi}

\begin{rmk}
{\rm
The affine groups can be defined in an alternative manner, in terms of the transformation laws of the $1$-forms $\a$ and $\b$, and then directly as subgroups of $K(1)$. The group $\Aff(1|1)$ is the subgroup of those $\Phi\in K(1)$ which satisfy $\Phi^*\b=F_\Phi\b$, with $F_\Phi$ a superfunction; restricting to $\pi(F_\Phi)>0$ we obtain the subgroup $\Aff_+(1|1)$. 
}
\end{rmk}
 
Acting by contactomorphisms on $\SII$, $\Aff_+(1|1)$ preserves the
orientation of the underlying circle, see Remark \ref{orientation}.
Moreover, two points on the supercircle $t_1$ and~$t_2$ define an
 orientation given by the sign of $x_2-x_1$ (in the chosen affine coordinate system). Hence, the action of $\Aff_+(1|1)$
 cannot be $2|1$-transitive, but for all couples $s$ and $t$ defining
 the same orientation there exists a unique $h\in\Aff_+(1|1)$ such that
 $\widehat{h}(t)\stackrel{2|1}{=} s$. So, let us introduce
 $\widetilde{\Aff_+(1|1)}$ as the group generated by $\Aff_+(1|1)$ and the orientation-reversing
 transformation $r:(x,\xi)\mapsto(-x,\xi)$.
\begin{lem}
The action of $\widetilde{\Aff_+(1|1)}$ on $\bbR^{1|1}\subset\SII$ is simply
 $2|1$-transitive.
\end{lem}
\begin{proof}
Let $a_1=(0,0)$, $a_2=(1,\zeta)$ and $t_1,t_2$ be two distinct 
points of $\SII$, with $x_1<x_2$, a condition which can always been satisfied, using the transformation $r$, if necessary. We look for
 $h\in\Aff_+(1|1)$ such that 
$\widehat{h}((t_1,t_2))\stackrel{2|1}{=}(a_1,a_2)$. We thus have to
 solve the system: 
$a^2x_1+a^2b-a^2\b\xi_1=0$, $a\b+a\xi_1=0$ and $a^2x_2+a^2b-a^2\b\xi_2=1$. In doing so, we obtain $\b=-\xi_1$, $b=- x_1$ and
 $a^2=[t_1,t_2]^{-1}$, see (\ref{I_euclidien}). This entails that $h$ is 
uniquely determined and 
$\widehat{h}(t_3)=\left(\frac{[t_1,t_3]}{[t_1,t_2]},\frac{\{t_1,t_3\}}{[t_1,t_2]^\half}
\right)$ for any point $t_3$ of $\SII$. Here, $p_1$ is the projection to the odd component, hence $p_1(\widehat{h}(t_2))$ is given by $\frac{\{t_1,t_2\}}{[t_1,t_2]^\half}$. The function $p_1\circ\widehat{h}$ is thus surjective from $[t]$ onto $\bbR^{0|1}$ and Proposition \ref{pro_pq-transitive} applies, proving the simply $2|1$-transitivity of the action of
 $\widetilde{\Aff_+(1|1)}$.
\end{proof}
Now, the action of $\widetilde{\Aff_+(1|1)}$ on $\SII$ satisfies all assumptions of Theorem \ref{generalinvariant}, and then, restricting ourselves to $x_1<x_2$, we obtain affine invariants with all properties stated in Theorem~\ref{generalinvariant}. 
\begin{pro}\label{AffineThm}
Choosing $\mathsf{a}$, the class of a couple $a=((0,0),(1,\zeta))$ for the relation~$\stackrel{2|1}{=}$, Theorem \ref{generalinvariant} gives rise to a characteristic affine
invariant consisting of the following three-point couple of superfunctions, defined, for $x_1<x_2$, by 
\begin{equation}\label{I_affine}
I_\mathsf{a}(t_1,t_2,t_3)=\left([t_1,t_2,t_3]\, 
,\,\{t_1,t_2,t_3\}\right)=\left(\frac{[t_1,t_3]}{[t_1,t_2]},\frac{\{t_1,t_3\}}{[t_1,t_2]^\half}\right).
\label{affineinvariant}
\end{equation} 
We, likewise, have a two-point odd invariant, defined, for $x_1<x_2$, by
\begin{equation}\label{J_affine}
J_\mathsf{a}(t_1,t_2)=\frac{\{t_1,t_2\}}{[t_1,t_2]^\half},
\label{Jaffineinvariant}
\end{equation} 
which is fundamental in that it generates all other two-point
 invariants.
\end{pro}

\begin{proof}
The action of $\widetilde{\Aff_+(1|1)}$ being simply $2|1$-transitive,
 we can apply Theorem \ref{generalinvariant}. Let $t=(t_1,t_2)$ be a couple; if
 $x_1<x_2$, we obtain, resorting to the proof of the last lemma,
 $I_\mathsf{a}(t_1,t_2,t_3)=\left(\frac{[t_1,t_3]}{[t_1,t_2]},\frac{\{t_1,t_3\}}{[t_1,t_2]^\half}
 \right)$ and $J_\mathsf{a}(t_1,t_2)=\frac{\{t_1,t_2\}}{[t_1,t_2]^\half}$. For
 $x_1<x_2$, $I_\mathsf{a}$ and $J_\mathsf{a}$ are invariants (with all
 properties given in Theorem \ref{generalinvariant}) of the subgroup of
 $\widetilde{\Aff_+(1|1)}$ preserving the condition $x_1<x_2$, i.e.,
 $\Aff_+(1|1)$.  
\end{proof}

For $x_2<x_1$, we can easily show that $I_\mathsf{a}$ and
 $J_\mathsf{a}$ are simply obtained by exchanging~$t_1$ and $t_2$. 

\begin{rmk}
{\rm
The invariants $I_\mathsf{a}$ and $J_\mathsf{a}$ depend on $\mathsf{a}$; for another class, $\mathsf{b}$, of a couple of points, we have, following the third assertion of Theorem \ref{generalinvariant}, $I_\mathsf{b}=\widehat{g}\circ I_\mathsf{a}$ and
$J_\mathsf{b}=\widehat{g}\circ J_\mathsf{a}$ iff $p_1(b_1)=0$. For $p_1(b_1)\neq 0$, $I_\mathsf{b}$ and $J_\mathsf{b}$ depend on $I_\mathsf{a}$ and $J_\mathsf{a}$ in a more involved way.
}
\end{rmk}

\begin{rmk}
{\rm
We can rewrite the odd three-point invariant, $p_1(I_\mathsf{a})$, as 
$\{t_1,t_2,t_3\}=[t_1,t_2,t_3]^\half\frac{\{t_1,t_3\}}{[t_1,t_3]^\half}$,
showing that it is a function of the odd two-point invariant function,
$J_\mathsf{a}$, and of the even three-point invariant, $p_0(I_\mathsf{a})$. Hence, every affine
three-point invariant function is a function of $J_\mathsf{a}$ and $p_0(I_\mathsf{a})$.
}
\end{rmk}

\begin{cor}\label{evenaff}
The even part, $p_0(I_\mathsf{a})$, of $I_\mathsf{a}$ is invariant under $\Aff(1|1)$, and
 characterizes this subgroup of $K(1)$, namely if $\Phi\in K(1)$ satisfies
 $\Phi^*[t_1,t_2,t_3]=[t_1,t_2,t_3]$, then $\Phi=\widehat{h}$ for some
 $h\in\Aff(1|1)$.
\end{cor}
The proof is identical to that of Corollary \ref{eveneucl}, in the
 Euclidean case.

%%%%%%%%%%%%%%%%%%%%%%%%%%%%%%%%%%%%%%%%%%%%%%%%%%%%%%%%%%%%
\subsection{Projective invariants}
%%%%%%%%%%%%%%%%%%%%%%%%%%%%%%%%%%%%%%%%%%%%%%%%%%%%%%%%%%%%

Once more, we will follow the previous method, and derive the super
 cross-ratio 
as the even part of the $\SpO_+(2|1)$-invariant given by Theorem
 \ref{generalinvariant}.

We begin by the introduction of an orientation index, $\ord$, on the oriented circle, defined on triples of distinct points by 
 \begin{eqnarray}\label{ord}
 \ord(x_1,x_2,x_3)&=& +1 \text{ if } x_2\in[x_1,x_3]\\\nonumber
 &=&-1  \text{ if } x_2\in[x_3,x_1].
 \end{eqnarray} 
It is uniquely preserved by
orientation-preserving diffeomorphisms of the circle and enjoys the property:
$\ord(\sigma(x_1),\sigma(x_2),\sigma(x_3))=\ve(\sigma)\ord(x_1,x_2,x_3)$ for
 any permutation~$\sigma$ whose parity is denoted by $\ve(\sigma)$, see \cite{BG}.
This index, $\ord$, can be extended to triples of points of the supercircle by  $\ord(t_1,t_2,t_3)=\ord(x_1,x_2,x_3)$. 

As $\SpO_+(2|1)$ acts by contactomorphisms on $\SII$, it preserves the
 orientation of the underlying circle, see Remark \ref{orientation}.
Hence, the action of $\SpO_+(2|1)$ cannot be $3|2$-transitive, a triple of distinct points defining an orientation. However, if $s$ and $t$ are two triples 
 defining the same orientation, there exist exactly two elements
 $h_\pm\in\SpO_+(2|1)$ such that $\widehat{h_\pm}(t)\stackrel{3|2}{=} s$. So, let
 us introduce $\widetilde{\SpO_+(2|1)}$, the group generated by
 $\SpO_+(2|1)$ already considered, and the orientation-reversing transformation $r:(x,\xi)\mapsto(-x,\xi)$.

\begin{lem}\label{SpO_transitive}
\begin{enumerate}
\item The action of $\widetilde{\SpO_+(2|1)}$ on $\SII$ is $3|2$-transitive. 
\item Moreover, let $\mathsf{p}$ be the class of $p=((\infty,0),(0,0),(1,\zeta))$ for the relation $\stackrel{3|2}{=}$, then for any
 triple $t$, there exist exactly two elements of
 $\widetilde{\SpO_+(2|1)}$, $k_+,k_-$, such that $\widehat{k}_\pm(t)\stackrel{3\vert 2}{=}p$,
 and $\widehat{k_-}=\iota\circ\widehat{k_+}$, with $\iota
 :(x,\xi)\mapsto (x,-\xi)$.
\end{enumerate} 
\end{lem}
\begin{proof}
Assume first that the triple $t=(t_1,t_2,t_3)$ be such that $x_1<x_2<x_3$, even if it means to apply $r$ and an element of $\SpO_+(2|1)$ inducing a cyclic permutation on~$t$. Then, Proposition~\ref{AffineThm} insures that 
there exists a unique $g\in\Aff_+(1|1)$ such that:
 $\widehat{g}(t_2)=(0,0)$, and 
$\widehat{g}(t_3)=(1,\zeta')$, with 
$\zeta'=\frac{\{t_2,t_3\}}{[t_2,t_3]^\half}$. Since $g\in\SpO_+(2|1)$,
 we just have to determine all $h\in\SpO_+(2|1)$ such that
 $\widehat{h}(0,0)=(0,0)$, 
$p_0(\widehat{h}(1,\zeta'))=1$, and
 $\widehat{h}(\widehat{g}(t_1))=(\infty,0)$, implying that 
$hg=k$ are the sought transformations such that $\widehat{hg}(t)\stackrel{3|2}{=}p$.

%\goodbreak

As $h$ is an element of $\SpO_+(2|1)$, $\widehat{h}$ is 
of the form 
$\widehat{h}(x,\xi)=\left(\frac{ax+b+\g\xi}{cx+d+\d\xi},\frac{\a
 x+\b+e\xi}{cx+d+\d\xi}\right)$, with the relations 
(\ref{ad-bc=1}) to (\ref{betae}). Since 
$\widehat{h}(0,0)=(0,0)$, we have $b=\beta=0$, and the relations become
 $ad=1$, 
 $e^2=1$, $\a e=a\d$ and $\g=0$; now $e=1$ since we restrict us to
 special transformations, i.e., of Berezinian $1$. The equation
 $\widehat{h}(\widehat{g}(t_1))=(\infty,0)$ gives
 $ac\frac{[t_2,t_1]}{[t_2,t_3]}+1+\alpha\frac{\{t_2,t_1\}}{[t_2,t_3]^\half}=0$ and
 $\a\frac{[t_2,t_1]}{[t_2,t_3]}+\frac{\{t_2,t_1\}}{[t_2,t_3]^\half} =0$, where we have used the fact that $\widehat{g}(t_1)=I_\mathsf{a}(t_2,t_3,t_1)$ as given by (\ref{I_affine}). Hence, we have 
 \begin{equation}\label{alpha_ac}
 \a = -\frac{\{t_1,t_2\}}{[t_2,t_3]^\half}\frac{[t_2,t_3]}{[t_1,t_2]} \qquad \text{and} \qquad 
ac=\frac{[t_2,t_3]}{[t_1,t_2]}.
\end{equation}
There is one extra equation to
 satisfy, namely $p_0(\widehat{h}(1,\zeta'))=1$; it yields 
explicitly $a^2=ac+1+\a \zeta'$, giving 
$a^2=\frac{[t_2,t_3]}{[t_1,t_2]}+1-\frac{\{t_1,t_2\}\{t_2,t_3\}}{[t_1,t_2]}$, since $\zeta'=J_\mathsf{a}(t_2,t_3)$, see (\ref{Jmj}), as given by~(\ref{J_affine}). We then get, with the help of the identity 
$[t_2,t_3]+[t_1,t_2]-\{t_1,t_2\}\{t_2,t_3\}=[t_1,t_3]$,
\begin{equation}\label{a2}
a^2=\frac{[t_1,t_3]}{[t_1,t_2]},
\end{equation}
 so $a$ is determined up to an overall sign.  We have proved that $h$ is therefore
 given by
 \begin{equation}\label{h_lemproj}
 \widehat{h}(x,\xi)=\left(\frac{a^2x}{acx+1+\a \xi},\frac{a(\a
 x+\xi)}{acx+1+\a \xi}\right),
 \end{equation}
 the sign of $a\neq0$ remaining unspecified. This
 proves the existence and uniqueness of~$h_\pm$, as stated above.
 Moreover, $J_\mathsf{p}(t_1,t_2,t_3)=p_1(\widehat{h}(t_3))$ is a surjective
 function from~$[t]$ to~$\bbR^{0\vert 1}$, see (\ref{Jp}). Using Proposition \ref{pro_pq-transitive}, we conclude that the action of
 $\widetilde{\SpO_+(2|1)}$ is $3\vert 2$-transitive.
\end{proof}
Now, even if the action of $\widetilde{\SpO_+(2|1)}$ is not simply $3|2$-transitive, we can construct, following Theorem \ref{generalinvariant}, invariants in the
 same way as before, and restricting consideration to $\ord(t_1,t_2,t_3)=1$, we will end up with
 projective invariants.

\begin{pro}\label{ProjectiveThm}
Let $p=((\infty,0),(0,0),(1,\zeta))$ be a triple of points of $\SII$, and denote by $\mathsf{p}$ the class of $p$ for the relation $\stackrel{3|2}{=}$. Theorem \ref{generalinvariant} then yields the projective invariant
$I_\mathsf{p}(t_1,t_2,t_3,t_4)=\left([t_1,t_2,t_3,t_4]\, ,\pm\{t_1,t_2,t_3,t_4\}\right)$, given, if $\ord(t_1,t_2,t_3)=1$, by
\begin{eqnarray}
\label{evenprojinv}
[t_1,t_2,t_3,t_4]&=&\frac{[t_1,t_3][t_2,t_4]}{[t_2,t_3][t_1,t_4]}, \\[6pt] 
\{t_1,t_2,t_3,t_4\}&=&[t_1,t_2,t_3,t_4]^\half\;
 \frac{\{t_2,t_4\}[t_1,t_2]-\{t_1,t_2\}[t_2,t_4]}{([t_1,t_2][t_2,t_4][t_1,t_4])^\half},
\label{oddprojinv}
\end{eqnarray} 
which characterizes the group $\SpO_+(2|1)$ within the diffeomorphisms
 of $\SII$. 

\goodbreak

We also have an odd projective invariant, namely, if
 $\ord(t_1,t_2,t_3)=1$, 
\begin{equation}\label{Jp}
J_\mathsf{p}(t_1,t_2,t_3)=\pm
 \frac{\{t_2,t_3\}[t_1,t_2]-\{t_1,t_2\}[t_2,t_3]}{([t_1,t_2][t_2,t_3][t_1,t_3])^\half}.
\label{epsilon}
\end{equation}
which is fundamental in that it generates all other three-point
 invariants. 
\end{pro}

\begin{proof}
Assume that $\ord(t_1,t_2,t_3)=1$. Using Lemma \ref{SpO_transitive}, we know that there exist exactly two elements $k_+,k_-\in\SpO_+(2|1)$ such that $\widehat{k_\pm}(t)\stackrel{3|2}{=}p$. We set $I_\mathsf{p}(t_1,\ldots,t_4)=\widehat{k_\pm}(t_4)$ and $J_\mathsf{p}(t_1,t_2,t_3)=p_1(\widehat{k_\pm}(t_3))$, as suggested by Theorem \ref{generalinvariant}. Despite the non uniqueness of $k$, all conclusions of Theorem~\ref{generalinvariant} apply just as well, and the proofs are identical, except for $I_\mathsf{p}$ being a characteristic invariant.
The proof of Theorem~\ref{generalinvariant} shows that any bijection, $\Phi$, of the supercircle such that $\Phi^*I_\mathsf{p}=I_\mathsf{p}$, satisfies $\Phi(t_4)=\widehat{k_\pm}(t_4)$ for all $t_4$. We have to impose that~$\Phi$ be a diffeomorphism to obtain $\Phi=\widehat{k_+}$ or $\Phi=\widehat{k_-}$. 
  
It then remains to compute $\widehat{k_\pm}(t_4)$; using the proof of Lemma \ref{SpO_transitive} we will easily calculate $\widehat{k_\pm}=\widehat{h_\pm}\circ\widehat{g}$, for the specific case $x_1<x_2<x_3$. Starting with the even part of $\widehat{k_\pm}(t_4)$, we obtain,
%for $x_1<x_2<x_3$
see (\ref{h_lemproj}),
\begin{eqnarray*}
[t_1,t_2,t_3,t_4]
&=& \frac{a^2[t_2,t_3,t_4]}{ac[t_2,t_3,t_4]+1+\a\{t_2,t_3,t_4\}}\\[6pt]
&=&
\frac{[t_1,t_3][t_2,t_4]}{[t_1,t_2][t_2,t_3]\left(\displaystyle{\frac{[t_2,t_4]}{[t_1,t_2]}+1-\frac{\{t_1,t_2\}\{t_2,t_4\}}{[t_1,t_2]}}\right)},
\end{eqnarray*}
where we have used (\ref{alpha_ac}) and (\ref{a2}).
With the help of the identity 
$[t_2,t_4]+[t_1,t_2]-\{t_1,t_2\}\{t_2,t_4\}=[t_1,t_4]$, we find the
 announced result, viz., Equation~(\ref{evenprojinv}).

We then compute the odd part of $\widehat{h_\pm}(\widehat{g}(t_4))$, which is determined up to global sign governed by the sign of $a$ (see proof of Lemma \ref{SpO_transitive}). 
For $a>0$, we find, using (\ref{h_lemproj}),
\begin{eqnarray*}
\{t_1,t_2,t_3,t_4\}&=&\frac{a\left(\a 
[t_2,t_3,t_4]+\{t_2,t_3,t_4\}\right)}{ac[t_2,t_3,t_4]+1+\a
 \{t_2,t_3,t_4\}}\\%[6pt]
&=&\frac{([t_1,t_2][t_1,t_3])^\half\left(\displaystyle{-\frac{\{t_1,t_2\}}{[t_2,t_3]^\half}\frac{[t_2,t_4]}{[t_1,t_2]}+\frac{\{t_2,t_4\}}{[t_2,t_3]^\half}}\right)}{[t_1,t_4]}\\%[6pt]
&=&[t_1,t_2,t_3,t_4]^\half\left([t_1,t_2,t_4]^{-\half}\frac{\{t_2,t_4\}}{[t_2,t_4]^\half}-[t_4,t_2,t_1]^{-\half}\frac{\{t_1,t_2\}}{[t_1,t_2]^\half}\right),
\end{eqnarray*}
with the help of the equalities (\ref{alpha_ac}) and (\ref{a2}).
For $x_1<x_2<x_3$, we can write
\begin{equation}
\{t_1,t_2,t_3,t_4\}= [t_1,t_2,t_3,t_4]^\half
 \frac{\{t_2,t_4\}[t_1,t_2]-\{t_1,t_2\}[t_2,t_4]}{([t_1,t_2][t_2,t_4][t_1,t_4])^\half},
\end{equation}
 which is the announced result, viz., Equation (\ref{oddprojinv}).

\goodbreak

For the more general case $\ord(t_1,t_2,t_3)=1$, we still have to compute $\widehat{k_\pm}(t_4)$ for $x_3<x_1<x_2$ and $x_2<x_3<x_1$. Let us introduce the homography $\widehat{c}(x,\xi)=(\frac{x-1+\zeta\xi}{x},\frac{\zeta x-\xi}{x})$, which cyclically permutes $(0,0)$, $(\infty,0)$ and $(1,\zeta)$. Start with the case $x_3<x_1<x_2$; we can assume that $x_3<0<x_1<x_2$, even if it means to apply a translation, and then $\widehat{c}(x_1)<\widehat{c}(x_2)<\widehat{c}(x_3)$. As $I_\mathsf{p}$ is invariant under the action of $\SpO_+(2|1)$, we have $I_\mathsf{p}=\widehat{c}{\,}^*I_\mathsf{p}$, and using the above results, we deduce that $\widehat{k_\pm}(t_4)=\widehat{c}{\,}^*\left(\frac{[t_1,t_3][t_2,t_4]}{[t_2,t_3][t_1,t_4]}, 
[t_1,t_2,t_3,t_4]^\half\;\frac{\{t_2,t_4\}[t_1,t_2]-\{t_1,t_2\}[t_2,t_4]}{([t_1,t_2][t_2,t_4][t_1,t_4])^\half}\right)$. The Euclidean invariants are transformed by $\widehat{c}$ as follows  $\widehat{c}^*[t_i,t_j]=\frac{[t_i,t_j]}{x_i x_j}$ and $\widehat{c}^*\{t_i,t_j\}=\frac{t_i}{x_i}-\frac{t_j}{x_j}$, we then have $\widehat{k_\pm}(t_4)=\left(\frac{[t_1,t_3][t_2,t_4]}{[t_2,t_3][t_1,t_4]}, 
[t_1,t_2,t_3,t_4]^\half\;\frac{\{t_2,t_4\}[t_1,t_2]-\{t_1,t_2\}[t_2,t_4]}{([t_1,t_2][t_2,t_4][t_1,t_4])^\half}\right)$. The case $x_2<x_3<x_1$ is similar, except for the fact that we have to apply $\widehat{c^2}$ instead of $\widehat{c}$.
%This ends the proof.
\end{proof}

For $\ord(t_1,t_2,t_3)=-1$, the projective invariants $I_\mathsf{p}$ and $J_\mathsf{p}$ are simply given by the exchange of $t_1$ and $t_2$ in Formul{\ae} (\ref{evenprojinv}), (\ref{oddprojinv}), and (\ref{epsilon}). 

\begin{cor}\label{evenproj}
The cross-ratio (\ref{evenprojinv}) is invariant under $\SpO_+(2|1)$, and characterizes
 this subgroup of $K(1)$, namely if $\Phi\in K(1)$ satisfies
 $\Phi^*[t_1,t_2,t_3,t_4]=[t_1,t_2,t_3,t_4]$, then $\Phi=\widehat{h}$ for some
 $h\in\SpO_+(2|1)$.
\end{cor}
The proof is the same as in the Euclidean case, except for the fact
 that $\SpO_+(2|1)$ contains now the transformation $\iota:(x,\xi)\mapsto(x,-\xi)$.

\begin{rmk}
{\rm
Projective groups and projective invariants of the circle and of the supercircle share various properties. Like $\SpO_+(2|1)$ in the super case, the action of $\PSL(2,\bbR)$ preserves the orientation of the circle. The action of $\PSL(2,\bbR)$ on the circle is thus not simply $3$-transitive, in contradistinction to that of $\PGL(2,\bbR)$. The cross-ratio can also be defined following Theorem \ref{generalinvariant}, leading to the classical expression $[x_1,x_2,x_3,x_4]=\frac{(x_1-x_3)(x_2-x_4)}{(x_2-x_3)(x_1-x_4)}$, which is invariant under $\PSL(2,\bbR)$ only. The $\PGL(2,\bbR)$-invariant is given either by this last expression or by the same expression where $x_1$ and~$x_2$ have been exchanged, depending on $\ord(x_1,x_2,x_3)$. 
}
\end{rmk}

\begin{rmk}
{\rm
Again, the odd four-point invariant $p_1(I_\mathsf{p})$ is a function of the odd three-point invariant $J_\mathsf{p}$ and of the even four-point invariant $p_0(I_\mathsf{p})$. So every four-point invariant is a function of these two invariants.
}
\end{rmk}

%%%%%%%%%%%%%%%%%%%%%%%%%%%%%%%%%%%%%%%%%%%%%%%%%%%%%%%%%%%%
%%%%%%%%%%%%%%%%%%%%%%%%%%%%%%%%%%%%%%%%%%%%%%%%%%%%%%%%%%%%
\section{The Schwarzian derivative from the Cartan formula}
\label{CartanSection}
%%%%%%%%%%%%%%%%%%%%%%%%%%%%%%%%%%%%%%%%%%%%%%%%%%%%%%%%%%%%
%%%%%%%%%%%%%%%%%%%%%%%%%%%%%%%%%%%%%%%%%%%%%%%%%%%%%%%%%%%%
\label{proofcocycle}
This section provides the proof of Theorem \ref{thmcocycle}. We will 
begin by two preliminary lemmas and then give the proof for the 
Euclidean and affine cases, and, finally, for the projective one. 

%%%%%%%%%%%%%%%%%%%%%%%%%%%%%%%%%%%%%%%%%%%%%%%%%%%%%%%%%%%%
\subsection{Preparation}  
%%%%%%%%%%%%%%%%%%%%%%%%%%%%%%%%%%%%%%%%%%%%%%%%%%%%%%%%%%%%

Let us first recall the formula for the Taylor expansion of a
 smooth superfunction $f\in\Cinfty(\SII)$ as given in \cite{Lei,DM}, namely
\begin{eqnarray}
\label{Taylor}
\nonumber
f(t_2)-f(t_1)&=&\sum_{i=1}^{n}\frac{1}{i!}\left((x_2-x_1)^i\partial^i_x 
f(t_1)+i(\xi_2-\xi_1)(x_2-x_1)^{i-1}\partial^{i-1}_x\partial_\xi
 f(t_1)\right)\\
\nonumber
&&\qquad+O((x_2-x_1)^{n+1},(\xi_2-\xi_1)(x_2-x_1)^n) \\
\nonumber
&=& \sum_{i=1}^{n}\frac{1}{i!}\left([t_1,t_2]^i\partial^i_x 
f(t_1)+i\{t_1,t_2\}[t_1,t_2]^{i-1}\partial^{i-1}_x D f(t_1)\right)\\
&&\qquad+O((x_2-x_1)^{n+1},(\xi_2-\xi_1)(x_2-x_1)^n).
\end{eqnarray}

The following lemma linking discrete variations and forms, will enable
 us to write Taylor expansions in terms of the differential forms $\a$
 and $\b$. We will skip its straightforward proof.

\begin{lem}\label{lemdiscret}
Let $X\in\Vect(\SII)$, and $\phi_{\ve}$ the associated flow. Putting
 $t_2=\phi_\ve(t_1)$, we have
\begin{equation}
[t_1,t_2]=\left\langle \ve X,\a\right\rangle(t_1)+O(\ve^2), 
\quad\mbox{and}\quad\{t_1,t_2\}=\left\langle \ve
 X,\b\right\rangle(t_1)+O(\ve^2).
\end{equation}
\end{lem} 
The next result is of central importance in the subsequent proof of
 Theorem \ref{thmcocycle}.

\begin{lem}
Let $\Phi=(\varphi,\psi)\in K(1)$ be a contactomorphism of $\SII$, and
 let
$t_2=\phi_\ve(t_1)$, where $\phi_{\ve}$ is the flow of a vector field 
$X$, then
\begin{eqnarray}\label{devt_ordre_3_eq}
\nonumber
\frac{\Phi^*[t_1,t_2]}{E_\Phi(t_1)[t_1,t_2]}&=& 1+
 \frac{1}{2}\left([t_1,t_2]\frac{E_\Phi'}{E_\Phi}(t_1)+\{t_1,t_2\}\frac{DE_\Phi}{E_\Phi}(t_1)\right)
 \\[6pt]
&&+\left\langle  \ve X\otimes \ve X,\a^2\,\frac{A}{6E_\Phi}+\a\b\,
 \frac{B}{2E_\Phi}\right\rangle (t_1) +O(\ve^3)
\end{eqnarray}
where $A=\varphi'''+\psi\psi'''$ and $B=D\varphi''-\psi D\psi''$, 
$\a^2$ and $\a\b$ being as in (\ref{tensorproduct}).
\label{developpement_ordre_3}
\end{lem}
\begin{proof}
We have
$\Phi^*[t_1,t_2]=[\Phi(t_1),\Phi(t_2)]=\varphi(t_2)-\varphi(t_1)-(\psi(t_2)-\psi(t_1))\psi(t_1)$, by virtue of (\ref{I_euclidien}).
 Using Taylor's formula (\ref{Taylor}), we obtain
\begin{eqnarray*}
\Phi^*[t_1,t_2]&=&\{t_1,t_2\}(D\varphi-\psi D\psi)(t_1)
 +[t_1,t_2](\varphi'+\psi\psi')(t_1)\\\nonumber
&&+
 [t_1,t_2]\left(\frac{1}{2}[t_1,t_2](\varphi''+\psi\psi'')(t_1)+\{t_1,t_2\}(D\varphi'-\psi D\psi')(t_1)\right)\\\nonumber
&&+[t_1,t_2]\left(\frac{1}{6}[t_1,t_2]^2(\varphi'''+\psi\psi''')(t_1)+\frac{1}{2}\{t_1,t_2\}[t_1,t_2](D\varphi''-\psi
 D\psi'')(t_1)\right)\\
&&+O(\ve^4).
\end{eqnarray*}
Then, as $\Phi\in K(1)$, Proposition \ref{lemcontactomorphism} yields
 $D\varphi-\psi D\psi=0$, and $\varphi'+\psi\psi'=E_\Phi$.  This entails
 that $\varphi''+\psi\psi''=E_\Phi'$, and $D\varphi'-\psi
 D\psi'=\frac{1}{2}D E_\Phi$. Lemma \ref{lemdiscret} then leads to the result.
\end{proof}
At first order in $\ve$ we obtain simply:
 $\frac{\Phi^*[t_1,t_2]}{[t_1,t_2]}=\left[ E_\Phi+ \frac{1}{2}\langle \ve
 X,dE_\Phi\rangle\right](t_1) +O(\ve^2)
$.
%%%%%%%%%%%%%%%%%%%%%%%%%%%%%%%%%%%%%%%%%%%%%%%%%%%%%%%%%%%%
\subsection{Proof of Theorem \ref{thmcocycle}}
%%%%%%%%%%%%%%%%%%%%%%%%%%%%%%%%%%%%%%%%%%%%%%%%%%%%%%%%%%%%

%-----------------------------------------------------------
\subsubsection{Euclidean and affine $K(1)$-cocycles, $\cE,\cA$}\label{EA-cocycle-section}
%-----------------------------------------------------------

The Cartan formula (\ref{CartanFormula}) yields a privileged means to
 define the Schwarzian derivative via a Taylor expansion of the
 cross-ratio. Much in the same way, we will construct $1$-cocycles via the
 Euclidean and affine even invariants. Thanks to the last lemma, we have
\begin{equation}
\frac{\Phi^*[t_1,t_2]}{[t_1,t_2]}=E_\Phi(t_1)+O(\ve).
\label{CartanEuclid}
\end{equation}
Hence, $\cE:\Phi\mapsto\log (E_\Phi)$ is a $1$-cocycle of $K(1)$, with
 values in $\cF_0(\SII)$; this justifies~(\ref{Euclideancocycle}).
 Note that $\log(E_\Phi)$ is well-defined since the reduced function $\pi(E_\Phi)=\pi(D\psi)^2$, see~(\ref{quotient}), is positive.

For the affine even invariant (\ref{I_affine}), we have, putting $t_2=\phi_\ve(t_1)$ and $t_3=\phi_{2\ve}(t_1)$,
\begin{eqnarray}
\nonumber
\frac{\Phi^*[t_1,t_2,t_3]}{[t_1,t_2,t_3]}-1
&=&
\frac{1+\frac{1}{2}\left\langle 
2\ve X,d(\log
 E_\Phi)\right\rangle(t_1)+O(\ve^2)}{1+\frac{1}{2}\left\langle \ve X,d(\log{}E_\Phi)\right\rangle(t_1)+O(\ve^2)}-1\\[6pt]
&=&\frac{1}{2}\left\langle \ve X,d(\log
 E_\Phi)\right\rangle(t_1)+O(\ve^2).
 \label{CartanAffine}
\end{eqnarray}
This implies that $\cA:\Phi\mapsto d(\log E_\Phi)$ is a $1$-cocycle of
 the group $K(1)$ of contactomorphisms, with values in the space,
 $\Omega^1(\SII)$, of $1$-forms on $\SII$. Using the projection on
 half-densities $\cF_\frac{1}{2}(\SII)$ given by $\a^\half\la D,\,\cdot\,\ra$, see
 Proposition \ref{split}, we still obtain an affine $1$-cocycle: 
$\rA:\Phi\mapsto \a^\half\left\langle D,d(\log 
E_\Phi)\right\rangle=\frac{DE_\Phi}{E_\Phi}\,\a^\half$. The
justification of~(\ref{Affinecocycle}) and~(\ref{DAffinecocycle}) is complete.

%-----------------------------------------------------------
\subsubsection{The Schwarzian derivative, $\cS$}\label{S-cocycle-section}
%-----------------------------------------------------------

We will now resort, \textit{verbatim}, to the Cartan formula
 (\ref{CartanFormula}) in order to derive the expression of the Schwarzian
 derivative (\ref{Projectivecocycle}) of a diffeomorphism $\Phi\in{}K(1)$. This
 formula involves the cross-ratio $[t_1,t_2,t_3,t_4]$ of four close by
 points; we will, hence, posit $t_2=\phi_\ve(t_1),
 t_3=\phi_{2\ve}(t_1)$, and $t_4=\phi_{3\ve}(t_1)$, where $\phi_\ve=\Id+\ve{}X+O(\ve^2)$ is
 the flow of a vector field $X$ of $\SII$. 

Let us then expand in powers of $\ve$ the following expression: 
\begin{equation}\label{cr1}
\frac{\Phi^*[t_1,t_2,t_3,t_4]}{[t_1,t_2,t_3,t_4]}-1=\frac{\displaystyle{\frac{\Phi^*[t_1,t_3]}{E_\Phi(t_1)[t_1,t_3]}\frac{\Phi^*[t_2,t_4]}{E_\Phi(t_2)[t_2,t_4]}-\frac{\Phi^*[t_2,t_3]}{E_\Phi(t_2)[t_2,t_3]}\frac{\Phi^*[t_1,t_4]}{E_\Phi(t_1)[t_1,t_4]}}}{1+O(\ve)}.
\end{equation}
We note that the terms $\frac{\Phi^*[t_1,t_3]}{E_\Phi(t_1)[t_1,t_3]}$
 and $\frac{\Phi^*[t_1,t_4]}{E_\Phi(t_1)[t_1,t_4]}$, with base point
 $t_1$, are explicitly given by Lemma \ref{developpement_ordre_3}. The
 remaining terms, with base point $t_2$, will be computed separately,
 using again Equation (\ref{devt_ordre_3_eq}) and the Taylor formula
 (\ref{Taylor}), viz., 
\begin{equation}\label{devtaux}
f(t_2)=f(t_1)+[t_1,t_2] f'(t_1)+\{t_1,t_2\}Df(t_1)+O(\ve^2),
\end{equation}
for a superfunction $f\in\Cinfty(\SII)$.

We have 
\begin{eqnarray*}
\frac{\Phi^*[t_2,t_3]}{E_\Phi(t_2)[t_2,t_3]}&=& 1+
 \frac{1}{2}\left([t_2,t_3]\frac{E_\Phi'}{E_\Phi}(t_1)+\{t_2,t_3\}\frac{DE_\Phi}{E_\Phi}(t_1)
\right) \\[6pt]
&&+\frac{1}{2}[t_1,t_2]\left([t_2,t_3]\left(\frac{E_\Phi'}{E_\Phi}\right)'(t_1)+\{t_2,t_3\}\left(\frac{DE_\Phi}{E_\Phi}\right)'(t_1)\right)\\[6pt]
&&+\frac{1}{2}\{t_1,t_2\}\left([t_2,t_3]D\left(\frac{E_\Phi'}{E_\Phi}\right)(t_1)+\{t_2,t_3\}D\left(\frac{DE_\Phi}{E_\Phi}\right)(t_1)\right)\\[6pt]
&&+\left\langle\ve X\otimes\ve X,\a^2\,\frac{A}{6E_\Phi}+\a\b\,
 \frac{B}{2E_\Phi}\right\rangle (t_1) +O(\ve^3),
\end{eqnarray*}
where the terms $A$ and $B$ are defined in Lemma
 \ref{developpement_ordre_3}. The other term  $\frac{\Phi^*[t_2,t_4]}{E_\Phi(t_2)[t_2,t_4]}$
 is, likewise, obtained by replacing in the latter expression $t_3$ by
 $t_4$, and $\ve X$ by $2\ve X$. From Lemma~\ref{lemdiscret} and Taylor's
 formula (\ref{devtaux}), we get
 $[t_2,t_3]=\la\ve{}X,\a\ra(t_1)+O(\ve^2)$ and $\{t_2,t_3\}=\la\ve{}X,\b\ra(t_1)+O(\ve^2)$. In particular
 $\{t_2,t_3\}\{t_1,t_2\}$ is thus of third order in~$\ve$, since $\la\ve{}X,\b\ra$ is an odd superfunction. We finally have
\begin{eqnarray*}
\frac{\Phi^*[t_2,t_3]}{E_\Phi(t_2)[t_2,t_3]}&=& 1+\frac{1}{2} \left([t_2,t_3]\frac{E_\Phi'}{E_\Phi}(t_1)+\{t_2,t_3\}\frac{DE_\Phi}{E_\Phi}(t_1)\right)
 \\[6pt]
&&
+\frac{1}{2}\left\langle \ve X\otimes\ve X, \a^2
\left(\frac{E_\Phi'}{E_\Phi}\right)'+2\a\b\left(\frac{DE_\Phi}{E_\Phi}\right)'\right\rangle(t_1)
 \\[6pt]
&&
+\left\langle\ve X\otimes\ve X,\a^2\,\frac{A}{6E_\Phi}+\a\b\,
 \frac{B}{2E_\Phi}\right\rangle (t_1) +O(\ve^3).
\end{eqnarray*}
This formula and Lemma \ref{developpement_ordre_3} help us find the
contribution of the first order terms of each product in the numerator of
Equation (\ref{cr1}); this contribution is found as 
$
([t_1,t_3]+[t_2,t_4]-[t_2,t_3]-[t_1,t_4])\frac{E_\Phi'}{2E_\Phi}(t_1)+(\{t_1,t_3\}+\{t_2,t_4\}-\{t_2,t_3\}-\{t_1,t_4\})\frac{D
 E_\Phi}{2 E_\Phi}(t_1)
=(\xi_1-\xi_2)(\xi_3-\xi_4)\frac{E_\Phi'}{2E_\Phi}(t_1)
$, which is of third order in $\ve$, since
$\xi_3-\xi_4=\xi_1-\xi_2+O(\ve^2)$. The right-hand side of (\ref{cr1}) is of second order in $\ve$ and we now compute it. We find
\begin{eqnarray*}
\frac{\Phi^*[t_1,t_2,t_3,t_4]}{[t_1,t_2,t_3,t_4]}-1
&=&
\frac{1}{4}\left\langle \ve
 X,\a\frac{E_\Phi'}{E_\Phi}+\b\frac{DE_\Phi}{E_\Phi}\right\rangle^2(t_1)\\[6pt]
&&
+\half\left\langle \ve X\otimes \ve
 X,\a^2\left(\frac{E_\Phi'}{E_\Phi}\right)'
+2\a\b\left(\frac{DE_\Phi}{E_\Phi}\right)'\right\rangle(t_1)\\[6pt]
&& 
-2\left\langle \ve X\otimes \ve X,\a^2\,\frac{A}{6E_\Phi}+\a\b\,
 \frac{B}{2E_\Phi}\right\rangle(t_1)+O(\ve^3).
\end{eqnarray*}
Collecting the terms involving $\a^2$ and $\a\b$, we put the latter
 expression in a nicer form, namely
\begin{eqnarray}
\label{devt_cr_AB}
\nonumber
\frac{\Phi^*[t_1,t_2,t_3,t_4]}{[t_1,t_2,t_3,t_4]}-1&=&\left\langle \ve
 X\otimes\ve X, \a^2 
\left(\frac{E_\Phi''}{2E_\Phi}-\frac{A}{3E_\Phi}-\frac{1}{4}\left(\frac{E_\Phi'}{E_\Phi}\right)^2\right)\right\rangle(t_1)\\[6pt]
&&+\left\langle \ve X\otimes\ve X, 
\a\b\left(\frac{DE_\Phi'}{E_\Phi}-\frac{B}{E_\Phi}-\frac{E_\Phi'\,DE_\Phi}{2E_\Phi^2}\right)\right\rangle(t_1).
\end{eqnarray}
Since $D\varphi=\psi D\psi$ and $E_\Phi=(D\psi)^2$, see Proposition
 \ref{lemcontactomorphism}, we calculate the terms $A$ and $B$ whose
 expression is given in Lemma \ref{developpement_ordre_3}; we find
$A=\varphi'''+\psi\psi'''=(E_\Phi-\psi\psi')''+\psi\psi'''=E_\Phi''-\psi'\psi''$,
 together with $B=D\varphi''-\psi D\psi''=\psi'' D\psi+2\psi' D\psi'=
\frac{1}{2}D^3E_\Phi+\frac{1}{4}\frac{E_\Phi'\,DE_\Phi}{E_\Phi}$. 
Plugging these quantities into~(\ref{devt_cr_AB}), we obtain
\begin{eqnarray}
\label{devt-cr}
\nonumber 
\frac{\Phi^*[t_1,t_2,t_3,t_4]}{[t_1,t_2,t_3,t_4]}-1&=&\left\langle \ve 
X\otimes\ve X, \a^2 
\left(\frac{1}{6}\frac{E_\Phi''}{E_\Phi}-\frac{1}{4}\left(\frac{E_\Phi'}{E_\Phi}\right)^2+\frac{1}{3}\frac{\psi'\psi''}{E_\Phi}\right)\right\rangle(t_1)\\[6pt]&&
+\left\langle \ve X\otimes\ve X, 
\a\b\left(\frac{1}{2}\frac{DE_\Phi'}{E_\Phi}-\frac{3}{4}\frac{E_\Phi'\,DE_\Phi}{E_\Phi^2}\right)\right\rangle(t_1).
\end{eqnarray}
Upon defining
\begin{equation}\label{Stilde} 
\widetilde{S}(\Phi)=\frac{DE_\Phi'}{E_\Phi}-\frac{3}{2}\frac{E_\Phi'\,DE_\Phi}{E_\Phi^2},
\end{equation}
we find
$D\widetilde{S}(\Phi)=\frac{E_\Phi''}{E_\Phi}-\frac{3}{2}\left(\frac{E_\Phi'}{E_\Phi}\right)^2-\frac{1}{2}\frac{DE_\Phi'\,DE_\Phi}{E_\Phi^2}$.
We also have $DE_\Phi'\,DE_\Phi=-4\psi'\psi''E_\Phi$. Inserting the
 latter result into (\ref{devt-cr}) and using the Cartan formula
 (\ref{CartanFormula}) to define the Schwarzian derivative,
 $\cS(\Phi)$, of the contactomorphism $\Phi$, we obtain
\begin{equation}
\cS(\Phi)=\frac{1}{6}\a^2
 D\widetilde{S}(\Phi)+\frac{1}{2}\a\b\widetilde{S}(\Phi).
\label{Schwarzianderivative}
\end{equation}
Thus, $\cS$ defines a $1$-cocycle of $K(1)$ with values in the
 space, $\cQ(\SII)$, of quadratic differentials, cf. Subsection
 \ref{omega,q}. Using the projection onto the $\frac{3}{2}$-densities,
 $\cF_\frac{3}{2}(\SII)$, given by $\a^\half\la D,\,\cdot\,\ra$, see Proposition
 \ref{split},  we still obtain a projective $1$-cocycle of $K(1)$, viz., 
\begin{equation}\label{Radul}
\rS(\Phi)=\a^\half\left\langle D, \cS(\Phi) \right\rangle=\frac{1}{4} 
\left(\frac{DE_\Phi'}{E_\Phi}-\frac{3}{2}\frac{E_\Phi'\,DE_\Phi}{E_\Phi^2}\right)\a^{3/2}.
\end{equation}
This ends the proof of (\ref{DProjectivecocycle}). 

Equation (\ref{Projectivecocycle}) can now be deduced from
 (\ref{Schwarzianderivative}) and (\ref{Radul}). Indeed, using   
\begin{equation}
\label{LDalpha}
L_D\a=2\b,
\end{equation}
 we find $\a^\half L_D\rS(\Phi)=\frac{1}{4}\a^\half
 L_D(\widetilde{S}(\Phi)\a^{\frac{3}{2}})=\frac{3}{2}\cS(\Phi)$.

%-----------------------------------------------------------
\subsubsection{The kernels of the $K(1)$-cocycles
$\cE,\cA,\cS$}\label{KerCocycles-section}
%-----------------------------------------------------------

- The subgroup of those $\Phi\in{}K(1)$ such that $\cE(\Phi)=0$ is
 characterized by the equation $E_\Phi=1$, see (\ref{Euclideancocycle}).
 Writing $\Phi=(\varphi,\psi)$, 
and using (\ref{E}), we find $D\psi=\epsilon$, with $\epsilon^2=1$.
 This entails that $\psi(x,\xi)=\epsilon(\beta+\xi)$, with
 $\beta\in\bbR^{0|1}$. The constraint (\ref{contactconstraint}) then leads to
 $\varphi(x,\xi)=x+b-\b\xi$, with~$b\in\bbR$. This proves that
 $\ker(\cE)=\rE(1|1)$.

- The kernel of the $1$-cocycle $\cA$, given by (\ref{Affinecocycle}), is
 determined by the equation $E_\Phi=a^2$, with $a\in\bbR^*$. The kernel of $\rA$ is given by the same equation, hence is equal to the kernel of $\cA$. The same
 computation as before clearly leads to $\Phi(x,\xi)=(a^2x+a^2
 b-a^2\b\xi,a\b+a\xi)$. Hence, $\ker(\rA)=\ker(\cA)=\Aff(1|1)$.

- The kernels of the $1$-cocycles $\cS$ and $\rS$, given respectively
 by (\ref{Projectivecocycle}), and (\ref{DProjectivecocycle}), clearly
 coincide. Suffice it to determine $\ker(\rS)$. Let us consider $\Phi\in
 K(1)$, then its Schwarzian derivative (\ref{DProjectivecocycle}) reads
 alternatively 
\begin{equation}
\label{Sphibis}
\rS(\Phi)=-\half E_\Phi^\half\, D^3(E_\Phi^{-\half})\,\alpha^{3/2}.
\end{equation} 
Hence, $\rS(\Phi)=0$ iff $\partial_x D(E_\Phi^{-\half})=0$. As $\partial_x D \chi_0=0$ implies, for $\chi_0$ an even super\-function, $\chi_0=c'x +d'+\d' \xi$, where $(c',d',\d')\in\bbR^{2|1}$; we obtain
 $E_\Phi=(c'x +d'+\d' \xi)^{-2}$. Consider now $h\in\SpO_+(2|1)$, whose action
 is given by (\ref{homographie}), then $E_{\,\widehat{h}}=(cx +d+\d
 \xi)^{-2}$. We thus have $E_\Phi=E_{\,\widehat{h}}$ for some
 $h\in\SpO_+(2|1)$, so that $\Phi= \widehat{h}\circ\widehat{g}$ with $g\in\rE(1|1)$ in
 view of the above result; this implies that $\Phi\in\SpO_+(2|1)$. The
 conclusion, $\ker(\rS)=\SpO_+(2|1)$, easily follows.
 
  The proof of Theorem \ref{thmcocycle} is complete.

%%%%%%%%%%%%%%%%%%%%%%%%%%%%%%%%%%%%%%%%%%%%%%%%%%%%%%%%%%%%
\subsection{Proof of Proposition \ref{cArS}}
%%%%%%%%%%%%%%%%%%%%%%%%%%%%%%%%%%%%%%%%%%%%%%%%%%%%%%%%%%%%
With the help of (\ref{df}), the affine cocycle $\cA$, given by
 (\ref{Affinecocycle}), can be recast into the form
 $\cA(\Phi)=E_\Phi^{-1}dE_\Phi=\a E_\Phi^{-1}E_\Phi'+\b E_\Phi^{-1}DE_\Phi$. Using Equation (\ref{LDalpha}), we obtain $(\a^\half L_D)^2=\a L_{D^2}+\b L_D$.
 
 Straightforward calculation yields the expressions of $L_D\cA$,
 $L_{D^2}\cA$, and~$\cA^2$, so that
\begin{equation}
 (\a^\half L_D)^2 \cA(\Phi) -\half \cA(\Phi)^2=
 \a^2\left(D\widetilde{S}(\Phi)-\half\frac{DE_\Phi
 DE_\Phi'}{E_\Phi^2}\right)+2\a\b\widetilde{S}(\Phi).
\end{equation} 
This formula leads directly to (\ref{NiceFormula}), using $\la
 D,\a^2\ra=0$ and $\la D,\a\b\ra=\half \a$, together with the expressions~(\ref{Stilde})
 and (\ref{Radul}) for $\widetilde{S}$ and $\rS$.

%%%%%%%%%%%%%%%%%%%%%%%%%%%%%%%%%%%%%%%%%%%%%%%%%%%%%%%%%%%%
%%%%%%%%%%%%%%%%%%%%%%%%%%%%%%%%%%%%%%%%%%%%%%%%%%%%%%%%%%%%
\section{Super Euclidean, affine and projective invariants, and $K(N)$-cocycles for $\rS^{1|N}$}
\label{S1N}
%%%%%%%%%%%%%%%%%%%%%%%%%%%%%%%%%%%%%%%%%%%%%%%%%%%%%%%%%%%%
%%%%%%%%%%%%%%%%%%%%%%%%%%%%%%%%%%%%%%%%%%%%%%%%%%%%%%%%%%%%

The aim of this section is to extend to $\rS^{1|N}$ the previous constructions, namely those of the Euclidean, affine and projective invariants, of the Euclidean
 and affine cocycles, and of the Schwarzian derivative for $N=2$. For $N\geq
 3$, the cross-ratio is badly transformed by contactomorphisms, which
 prevents the construction of a Schwarzian derivative along the same
 lines as before (see Remark \ref{N=3_obstruction} below).
 
Let us define the notation used throughout this section. Except if
 otherwise stated, all indices $i,j$ of odd objects will run from $1$ to
 $N$, and Einstein's summation convention will be freely used.  The space
 of superfunctions $\cC^\infty(\rS^{1|N})$, defining $\rS^{1|N}$, is the superalgebra
 $\cC^\infty(\rS^{1})[\xi^1,\ldots,\xi^N]$ where the $\xi^i$ are odd indeterminates. It is
 topologically generated, as an algebra, by the coordinates $(x,\xi)$ with
 $\xi=(\xi^1,\ldots,\xi^N)$. The diffeomorphisms retain the form
 $\Phi=(\varphi,\psi)$, with $\psi=(\psi^1,\ldots,\psi^N)$ and
 $\varphi,\psi^j\in\cC^\infty(\rS^{1|N})$, such that $(\varphi,\psi)$ is a new
 coordinate system. Let $F,G\in\cC^\infty(\rS^{1|N})^N$, we denote their
 pairing with values in $\cC^\infty(\rS^{1|N})$ by
\begin{equation}\label{scalar-product}
F\cdot G = F_iG^i 
\end{equation}
where $F^i$ and $G^i$ are the $i$-th components of $F$ and $G$, and
 $F_i=\d_{ij} F^j$ (with the choice of an Euclidean signature).
The $\cC^\infty(\rS^{1|N})$-module $\Omega^1(\rS^{1|N})$ is generated
 by the $1$-forms
\begin{equation}\label{Nalphabeta}
\a=dx+\xi_i d\xi^i =dx+\xi\cdot d\xi \qquad \text{and}\qquad \b^i=d\xi^i,  
\end{equation}
with dual vectors $\partial_x$ and $D_i=\partial_{\xi_i}+\xi_i\partial_x$.
For $f\in\cC^\infty(\rS^{1|N})$ we therefore have
\begin{equation}
df=\a f'+\b^i D_i f,
\end{equation}
see (\ref{df}). We furthermore denote by $K(N)$ the group of
 contactomorphisms, $\Phi$, characterized by $\Phi^*\a=E_\Phi\,\a$ for some
 superfunction $E_\Phi$. Let $\Phi=(\varphi,\psi)\in K(N)$, then
 $\Phi^*\a=d\varphi + \psi\cdot d\psi= \a (\varphi'+\psi\cdot \psi')+
 \b^i(D_i\varphi-\psi\cdot D_i\psi)$. It follows that $\Phi\in K(N)$~iff
\begin{equation}
D_i\varphi-\psi \cdot D_i\psi=0,
\label{N,contactconstraint}
\end{equation} 
for all $i=1,\ldots, N$. The multiplier of $\Phi$ is then given by $E_\Phi=\varphi'+\psi\cdot \psi'$, i.e., by
\begin{equation}
E_\Phi=\frac{\Phi^*\a}{\a}=(D_i\psi)^2
\label{N,E}
\end{equation}
for any $i=1,\ldots,N$. The expression $(D_i\psi)^2$ stands for $D_i\psi\cdot
 D_i\psi$.
This has been first developed in the framework of super Riemann
 surfaces by Cohn~\cite{Coh}; we will nevertheless refer to work of Radul
 \cite{Rad}, whose geometric approach, in terms of contact structure, is
 closer to our viewpoint. See also \cite{GR}.
 \begin{pro}\label{Dortho}
Let $\Phi\in K(N)$, then 
\begin{equation}\label{Dipsi=orth}
D_jD_i\varphi+\psi\cdot D_jD_i\psi= D_i\psi\cdot D_j\psi=E_\Phi\delta_{ij}.
\end{equation}
 Hence $(E_\Phi^{-\half}D_i\psi)_{i=1,\ldots, N}$ is an ``orthonormal basis''
 for the pairing (\ref{scalar-product}) on
 $\Cinfty(\rS^{1|N})^N$.
 \end{pro}
\begin{proof}
As $\Phi\in K(N)$, we have $D_j
 D_i\varphi=D_j(\psi\cdot D_i \psi)=D_j\psi\cdot D_i \psi-\psi\cdot D_j
 D_i \psi$, in view of (\ref{N,contactconstraint}); by exchanging $i$ and $j$, we deduce $D_i\psi\cdot D_j \psi=0$
 if $i\neq j$. For $i=j$, the result is given by (\ref{N,E}) and the equality $E_\Phi=\varphi'+\psi\cdot\psi'$.
\end{proof}

%%%%%%%%%%%%%%%%%%%%%%%%%%%%%%%%%%%%%%%%%%%%%%%%%%%%%%%%%%%%
\subsection{Euclidean, affine and projective invariants}
%%%%%%%%%%%%%%%%%%%%%%%%%%%%%%%%%%%%%%%%%%%%%%%%%%%%%%%%%%%%

We now extend to $\rS^{1\vert N}$, where $N\geq2$, the content of
 Subsection \ref{SpO}. Now, $\a$ (\ref{Nalphabeta}) stems from the $1$-form on $\bbR^{2|N}$
 given by $\varpi=\frac{1}{2}(pdq-qdp+\theta_i d\theta^i)$, via  the formula
 $\varpi=\frac{1}{2}p^2\a$ ($p\neq 0$), 
expressed in affine coordinates $x=q/p$ and $\xi^i=\theta^i/p$. We 
define the orthosymplectic group \cite{GLS,Man}, $\SpO(2|N)$, as the 
supergroup whose $A$-points are all linear transformations of $\mathcal{O}_A^{2\vert N}$, see Subsection \ref{SpO},
\begin{equation}\label{matrix}
h=\left(\begin{array}{ccc} a&b&\g\\ c&d&\d\\ \a&\b&e \\ \end{array} 
\right)
\label{Northosymplectic}
\end{equation}
preserving the symplectic form $d\varpi$. 
If we demand that these linear transformations preserve the direction of $d\varpi$, only, we end up with the conformal supergroup $\mathrm{C}(2|N)$, see \cite{Man}. In the expression (\ref{matrix}), the entries $a,b,c,d$ are even elements,
 $\a,\b$ are odd column vectors of size $N$, while $\d,\g$ are odd row
 vectors of size $N$, and $e$ is an even matrix of size $N\times N$.
 Moreover, as $d\varpi$ is preserved, we have
\begin{eqnarray}
\label{N,ad-bc=1}
ad-bc-\a^t\b&=&1,\\
\label{N,e2=1}
e^t e+2\g^t\d&=&1,\\
\label{N,alphae}
\a^t e-a\d+c\g&=&0,\\
\label{N,betae}
\b^t e-b\d+d\g&=&0,
\end{eqnarray}
where the superscript $t$ denotes transposition. We easily find that
 $\SpO(2|N)$ also preserves $\varpi$. Again, since
 $\varpi=\frac{1}{2}p^2\a$, the orthosymplectic group acts by 
contacto\-morphisms, $\SpO(2|N)\rightarrow K(N)$, via the following
 projective action on $\rS^{1|N}$, namely
\begin{equation}
\widehat{h}(x,\xi)=\left( \frac{ax+b+\g \xi}{cx+d+\d\xi},\frac{\a 
x+\b+e\xi}{cx+d+\d\xi}\right),
\label{Nhomographie}
\end{equation}
where $h\in\SpO(2|N)$, and $\xi$ is understood as a column vector. 

\goodbreak

The kernel of this action is $\{\Id,-\Id\}$, hence the action is effective for the supergroup $\SpO(2|N)/\{\pm\Id\}=\PC(2|N)$ of conformal projective transformations. If $N$ is odd, this supergroup coincides with the special orthosymplectic group $\SpO_+(2|N)$, which is the subgroup of $\SpO(2|N)$ of Berezinian $1$.
We still can define Euclidean and affine subgroups of $\SpO(2|N)$, whose elements are
\begin{equation}
g=\left(
\begin{array}{ccc} 
a&a b&-a\b^t\\ 
0&a^{-1}&0\\ 
0&\b&1 \\ 
\end{array} 
\right),
\end{equation} 
where $(a,b,\b)\in\bbR^{2|N}$, $a>0$ defining $\Aff_+(1|N)$ and $a=1$ defining $\rE_+(1|N)$.
\begin{rmk}
{\rm
The group $\Aff_+(1|N)$ may be defined as the subgroup of those $\widehat{h}\in{}K(N)$ that preserve the direction of each $\beta_i$, namely $\widehat{h}^*\beta_i=\beta_i f_i$, for some super\-function $f_i$, with $i=1,\ldots, N$. Its subgroup $\rE_+(1|N)$ is characterized by exactly preserving $\a$. 
}
\end{rmk}

Let $t_1,t_2,t_3,t_4$ be four generic points of $\rS^{1|N}$.

\begin{thm}\label{N,generalinvariant}
We have three invariants, $I_\mathsf{e}$, $I_\mathsf{a}$ and $I_\mathsf{p}$, of the action of the Euclidean, affine and projective supergroups on the supercircle $\rS^{1\vert N}$.
\begin{itemize}
\item Euclidean invariant:
 $I_\mathsf{e}(t_1,t_2)=([t_1,t_2],\{t_1,t_2\})$ with
\begin{equation}
[t_1,t_2]=x_2-x_1-\xi_2\cdot\xi_1, \qquad\qquad \{t_1,t_2\}=\xi_2-\xi_1.
\end{equation}
\item Affine invariant: 
$I_\mathsf{a}(t_1,t_2,t_3)=([t_1,t_2,t_3],\{t_1,t_2,t_3\})$, where, if $x_1<x_2$,
\begin{equation}
[t_1,t_2,t_3]=\frac{[t_1,t_3]}{[t_1,t_2]}, \qquad \qquad
\{t_1,t_2,t_3\}=\frac{\{t_1,t_3\}}{[t_1,t_2]^\half}.
\end{equation}
\item Projective invariant: 
$I_\mathsf{p}(t_1,t_2,t_3,t_4)=([t_1,t_2,t_3,t_4],\rO(N).\{t_1,t_2,t_3,t_4\})$, where, if $\ord(t_1,t_2,t_3)=1$,
\begin{eqnarray}
\label{N,projectiveinvariant}
[t_1,t_2,t_3,t_4]&=&\frac{[t_1,t_3][t_2,t_4]}{[t_2,t_3][t_1,t_4]},
 \\[6pt] 
\{t_1,t_2,t_3,t_4\}&=&[t_1,t_2,t_3,t_4]^\half\;
 \frac{\{t_2,t_4\}[t_1,t_2]-\{t_1,t_2\}[t_2,t_4]}{([t_1,t_2][t_2,t_4][t_1,t_4])^\half}. 
\end{eqnarray} 
\end{itemize} 
The odd invariant, denoted by $\rO(N).\{t_1,t_2,t_3,t_4\}$, is the orbit of $\{t_1,t_2,t_3,t_4\}$ under the natural group action of $\rO(N)$. If a bijective transformation, $\Phi$, of $\rS^{1|N}$ leaves $I_\mathsf{e}$ (resp. $I_\mathsf{a}$) invariant, it can be identified with the action of an
element $h\in{}\rE_+(1|N)$ (resp. $h\in\Aff_+(1|N)$), i.e.,
$\Phi=\widehat{h}$. If $\Phi\in\Diff(\rS^{1|N})$ preserves $I_\mathsf{p}$, then $\Phi=\rho\circ\widehat{h}$, with $h\in\SpO(2|N)$ and $\rho(x,\xi)=(x,R\xi)$, $R\in\cC^\infty(\rS^{1|N},\rO(N))$. 
\end{thm} 

The proof of this Theorem can be carried out along the same lines as in the proof of Theorem \ref{thminvariant}, we will skip it and just provide some hints for it.
As in the case~$N=1$, we can show that the action of $\rE_+(1|N)$ is simply $1|1$-transitive, while that of $\widetilde{\Aff_+(1|N)}$ is simply $2|1$-transitive, on $\bbR^{1|N}\subset\rS^{1|N}$. Moreover, the action of~$\widetilde{\PC(2|N)}$ is $3|2$-transitive on $\rS^{1|N}$ and satisfies the following property: for any triple~$t$, and $g,h\in\widetilde{\PC(2|N)}$, $\widehat{g}(t)\stackrel{3|2}{=}p\stackrel{3|2}{=}\widehat{h}(t)$ is equivalent to $\widehat{g}=\widehat{k}\circ\widehat{h}$, with $\widehat{k}(x,\xi)=(x,e\xi)$, $e\in\rO(N)$. As in Section \ref{InvariantSection}, the tilde denotes the extension of the group by the involution $\iota:(x,\xi)\mapsto (-x,\xi)$. We can now apply Theorem \ref{generalinvariant} and the claims of Theorem \ref{N,generalinvariant} follow.

\begin{rmk}\label{Nevengroupinv}
{\rm
For $N=1$, the Corollaries \ref{eveneucl}, \ref{evenaff} and \ref{evenproj} have been obtained thanks to Lemma \ref{even-all}. They cannot be prolonged for $N>1$ as there exists no such lemma in this case. However, the supergroup preserving each even invariant is included in the kernel of the associated $K(N)$-cocycles, and for $N=2$, see Remark \ref{kernelRemark} and Theorem~\ref{thm2Sch}, one can easily check the converse inclusion. So, for $N=2$, the preserving supergroups of the even part of $I_\mathsf{e}$, $I_\mathsf{a}$ and $I_\mathsf{p}$ are respectively $\EO(1|2)/\{\pm\Id\}$, $\AO(1|2)/\{\pm\Id\}$ and $\PC(2|2)$.
}
\end{rmk}

%%%%%%%%%%%%%%%%%%%%%%%%%%%%%%%%%%%%%%%%%%%%%%%%%%%%%%%%%%%%
\subsection{Associated cocycles from the Cartan formula}
%%%%%%%%%%%%%%%%%%%%%%%%%%%%%%%%%%%%%%%%%%%%%%%%%%%%%%%%%%%%

The following calculation will rely on Proposition \ref{Dortho}, and
 on the relation $[D_i,D_j]=D_iD_j+D_jD_i=2\d_{ij}\partial_x$, for
 $i,j=1,\ldots, N$, which results from a direct calculation. 
As in the case $N=1$, we need a lemma giving the third-order Taylor
 expansion of~$\Phi^*[t_1,t_2]$. To that end, we will be using the
 notation: 
\begin{equation}\label{bibj}
\b^i\b^j=\half(\b^i\otimes\b^j-\b^j\otimes\b^i),
\end{equation}
and
 $\b^i\b^j\b^k=\frac{1}{6}(\sum_{\sigma\in\mathfrak{S}_3}\ve(\sigma)\b^{\sigma(i)}\otimes\b^{\sigma(j)}\otimes\b^{\sigma(k)})$, i.e., the
 symmetrized tensor product of odd elements; see \cite{DM}.

\begin{lem}\label{N,taylordevt}
Let $\Phi=(\varphi,\psi)\in K(N)$, and 
$t_2=\phi_\ve(t_1)$, with $\phi_{\ve}$ the flow of a vector field~$X$, and $t_1$ a point of $\rS^{1|N}$;
 we then have
\begin{eqnarray}\label{Ndvtordre3}
\Phi^*[t_1,t_2]&=&[t_1,t_2]E_\Phi(t_1)\left(
 1+\half[t_1,t_2]\frac{E_\Phi'}{E_\Phi}(t_1)+\half\{t_1,t_2\}^i \frac{D_i E_\Phi}{E_\Phi}(t_1)
 \right)  \\[6pt]\nonumber 
  && +[t_1,t_2]\left\langle \ve X\otimes \ve X,
\a^2\frac{A}{6}+\a\b^i \frac{B_i}{2}+\b^i\b^j \frac{C_{ij}}{2}\right\rangle(t_1)
 \\[6pt] \nonumber 
&&+\frac{1}{6}\left\langle \ve X\otimes \ve X\otimes \ve X,\b^i\b^j\b^k \left[ D_k
 D_j D_i \varphi-\psi\cdot D_k D_j D_i \psi\right]
 \right\rangle(t_1)+O(\ve^4),
\label{Ndevt_ordre3}
\end{eqnarray}
where $A=\varphi'''+\psi\cdot\psi'''$, $B_i=D_i\varphi''-\psi\cdot
 D_i\psi''$ and $C_{ij}=D_j D_i\varphi'+\psi\cdot D_j D_i\psi'$.
\label{Ndeveloppement_ordre_3}
\end{lem}
\begin{proof}
By definition we have: 
$\Phi^*[t_1,t_2]=\varphi(t_2)-\varphi(t_1)-(\psi(t_2)-\psi(t_1))\cdot\psi(t_1)$.
 Using the formula (\ref{Taylor}), trivially extended to the case
 $N\geq2$, we obtain
 
 \goodbreak
 
\begin{eqnarray}
\label{eqNdevt_ordre3}
\nonumber
\Phi^*[t_1,t_2]&=&[t_1,t_2]\left[
 E_\Phi+\half[t_1,t_2](\varphi''+\psi\cdot\psi'')+\half\{t_1,t_2\}^i D_i(\varphi'+\psi\cdot\psi')\right]
  (t_1)\\\nonumber
&&+\half \{t_1,t_2\}^i\{t_1,t_2\}^j\left[  D_j D_i\varphi+\psi\cdot D_j
 D_i\psi\right] (t_1)\\\nonumber
&&+[t_1,t_2]\left[\frac{1}{6}[t_1,t_2]^2(\varphi'''+\psi\cdot\psi''')+\half\{t_1,t_2\}^i[t_1,t_2](D_i\varphi''-\psi\cdot
 D_i\psi'')\right](t_1) \\\nonumber
&& +\half [t_1,t_2]\{t_1,t_2\}^i\{t_1,t_2\}^j\left[ D_j D_i\varphi'+\psi\cdot
 D_j D_i\psi')\right](t_1) \\ \nonumber 
&&+\frac{1}{6}\{t_1,t_2\}^i\{t_1,t_2\}^j\{t_1,t_2\}^k\left[ D_k D_j
 D_i\varphi-\psi\cdot D_k D_j D_i\psi\right] (t_1)\\[6pt]
&& +O(\ve^4).
\end{eqnarray}
 The coefficient of $\{t_1,t_2\}^i\{t_1,t_2\}^j$ on the
 second line of (\ref{eqNdevt_ordre3}) vanishes if $i\neq j$, using~(\ref{Dipsi=orth}).
 The analog of Lemma \ref{lemdiscret} holds true, namely
 $[t_1,t_2]=\left\langle \ve X,\a\right\rangle+O(\ve^2)$, and
 $\{t_1,t_2\}^i=\left\langle \ve X,\b^i\right\rangle+O(\ve^2)$. Hence, we are done.
\end{proof}

\begin{rmk}\label{N=3_obstruction}
{\rm
The last term in (\ref{eqNdevt_ordre3}) definitely does not vanish in the case $N\geq 3$, implying that $\Phi^*[t_1,t_2]$ is not proportional to $[t_1,t_2]$ at third order in $\ve$. This entails that the Cartan formula fails to provide an expression of the Schwarzian derivative for~$N\geq 3$.
}
\end{rmk}

%-----------------------------------------------------------
\subsubsection{Euclidean and affine $K(N)$-cocycles}
%-----------------------------------------------------------

Up to the second order in $\ve$, $\Phi^*[t_1,t_2]$ is proportional to
 $[t_1,t_2]$; this enables us to obtain $1$-cocycles from Euclidean and
 affine invariants, as was done in Subsection \ref{EA-cocycle-section}.
\begin{thm}\label{NcocycleEA}
From the Euclidean and affine even invariants, we construct the two
 following $K(N)$ nontrivial $1$-cocycles:
\begin{itemize}
\item The Euclidean cocycle $\cE:K(N)\rightarrow \cF_0(\rS^{1\vert N})$
 :
\begin{equation}
\cE(\Phi)=\log(E_\Phi)=\log(D_i\psi)^2,
\end{equation}
where the equality holds for any $i=1,\ldots,N$.
\item The affine cocycle $\cA:K(N)\rightarrow \Omega^1(\rS^{1\vert
 N})$:
\begin{equation}
\cA(\Phi)=d\cE(\Phi)=\frac{dE_\Phi}{E_\Phi}.
\end{equation}
\end{itemize}
\end{thm}
The proof is the same as in the case $N=1$, it relies on Lemma
 \ref{Ndeveloppement_ordre_3}. 
\begin{rmk}\label{ProjectionRemark}
{\rm
The directions of the individual vector fields $D_i$ are no longer preserved by
 the contactomorphisms; only that of
 $D_1\otimes\cdots\otimes D_N$ is preserved. Hence, the projection of $\cA$ on $D_i$ is
 no longer a $K(N)$-cocycle. 
}
\end{rmk}

Let us introduce $\AO(1|N)$, the ortho-affine subgroup of $\SpO(2|N)$ whose elements are 
\begin{equation}
g=\left(
\begin{array}{ccc} 
a&a b&-a\b^t\\ 
0&a^{-1}&0\\ 
0&\b&e \\ 
\end{array} 
\right)
\end{equation} 
where $(a,b,\b)\in\bbR^{2|N}$, $e\in \rO(N)$, and restricting us to $a=\pm1$, we obtain the ortho-Euclidean subgroup $\EO(1|N)$. Since the action of $\SpO(2|N)$ on the supercircle has a kernel equal to $\{\pm\Id\}$, the same holds for its above introduced subgroups. 

\begin{rmk}\label{kernelRemark}
{\rm
For $N=2$, a direct computation shows that the kernel of the two cocycles $\cE$ and $\cA$, are,
respectively, $\EO(1|2)/\{\pm\Id\}$ and $\AO(1|2)/\{\pm\Id\}$. This groups are also the groups  preserving the even part of $I_\mathsf{e}$ and $I_\mathsf{a}$, see Remark \ref{Nevengroupinv}. But for $N\geq 3$, this is no longer the case, i.e., the subgroup of $K(N)$ preserving the even invariant and the kernel of the
 associated cocycle are no longer the same defining groups. For example, if
 $N=3$, the contactomorphism $\Phi=(\varphi,\psi)$, with
 $\varphi(x,\xi)=x+\xi_1\xi_2\xi_3 \l$ and $\psi(x,\xi)=\xi-(\xi_2\xi_3, \xi_3\xi_1,
 \xi_1\xi_2)\l$, where $\l\in\bbR^{0|1}$, does not preserve $p_0(I_\mathsf{e})$ although
 $E(\Phi)=1$. Moreover, $\Phi$ is not even an homography. 
 %The last term in~(\ref{eqNdevt_ordre3}) helps us understand this phenomenon: it does not vanish whenever $N\geq 3$ and is not proportional to $[t_1,t_2]$.  
}
\end{rmk}

%-----------------------------------------------------------
\subsubsection{The Schwarzian $K(2)$-cocycle}
%-----------------------------------------------------------

For $N=2$, the expression (\ref{N,taylordevt}) of $\Phi^*[t_1,t_2]$ is proportional to $[t_1,t_2]$. This enables us to use the Cartan formula  to define the projective $1$-cocycle, $\cS$, from the cross-ratio~(\ref{N,projectiveinvariant}). By construction, our
 projective $1$-cocycle will take its values in the $K(2)$-module of
 quadratic differentials, $\cQ(\rS^{1\vert2})$, generated by $\a^2$, $\a\b^1$,
 $\a\b^2$ and $\b^1\b^2$, where $\a^2$ and $\a\b^i$ are as in
 (\ref{tensorproduct}), and $\b^1\b^2$ as in (\ref{bibj}). One can check that the
 linear mapping 
\begin{equation}\label{aD1D2}
\a \la D_2\otimes{}D_1,.\ra: \cQ(\rS^{1\vert2})\to\cF_1(\rS^{1\vert2})
\end{equation}
intertwines the natural action of $K(2)$, see Remark
 \ref{ProjectionRemark}.

Now, the Schwarzian derivative given by Radul \cite{Rad}, or Cohn \cite{Coh}, for $N=2$, has again coefficients in tensor densities.
 Projecting the $1$-cocycle, $\cS$, via (\ref{aD1D2}), we will readily recover
 Radul's and Cohn's Schwarzian derivative.
\begin{thm}\label{thm2Sch}
From the cross-ratio (\ref{N,projectiveinvariant}), we deduce, via the
 Cartan formula (\ref{CartanFormula}), the following projective $1$-cocycle~$\cS:K(2)\rightarrow \cQ(\rS^{1\vert 2})$, which reads 
\begin{equation}\label{2Sch}
\cS=\frac{1}{6}\alpha^2\;
\left(D_1D_2\rS_{12}+\frac{1}{2}\rS_{12}^2\right)
+\frac{1}{2}\alpha(\b^1D_2+\b^2D_1)\rS_{12}+\beta^1\beta^2\; \rS_{12},  
\end{equation}
where we have put $\rS_{12}=2\,\rS\,\a^{-1}$, see (\ref{2SD}).

Moreover, using the projection (\ref{aD1D2}) of the quadratic differentials on $1$-densities, we
 obtain the Schwarzian derivative $\rS:K(2)\rightarrow \cF_1(\rS^{1\vert 2})$
 given by
\begin{equation}\label{2RSch}
\rS(\Phi)=\left( \frac{D_2
 D_1E_\Phi}{E_\Phi}-\frac{3}{2}\frac{D_2E_\Phi D_1E_\Phi}{ E_\Phi^2}\right) \a.
\label{2SD}
\end{equation}
These two $1$-cocycles are nontrivial; their kernels coincide and are isomorphic to $\PC(2|2)$.
\end{thm}
\begin{proof}
The formula of the cross-ratio being similar to that of the case $N=1$,
 we have to compute, again, the expression (\ref{cr1}), the term
 $\frac{\Phi^*[t_1,t_2]}{E_\Phi(t_1)[t_1,t_2]}$ being now given by Lemma~\ref{Ndeveloppement_ordre_3}. Straightforward calculation, essentially the same
 as in Subsection \ref{S-cocycle-section}, leads to
\begin{eqnarray*}
\frac{\Phi^*[t_2,t_3]}{E_\Phi(t_2)[t_2,t_3]}&=& 1+
 \frac{1}{2}\left([t_2,t_3]\frac{E_\Phi'}{E_\Phi}(t_1)+\{t_2,t_3\}^i\frac{D_iE_\Phi}{E_\Phi}(t_1)\right)
 \\[6pt]
&+\displaystyle{\frac{1}{2}}&\left\langle \ve X\otimes \ve X, \a^2
 \left(\frac{E_\Phi'}{E_\Phi}\right)'+2\a\b^i\left(\frac{D_iE_\Phi}{E_\Phi}\right)'+\b^i\b^j
 D_j\left(\frac{D_iE_\Phi}{E_\Phi}\right)\right\rangle(t_1) \\[6pt]
&+&\left\langle  \ve X\otimes \ve X,\a^2\,\frac{A}{6E_\Phi}+\a\b^i\,
 \frac{B_i}{2E_\Phi}+\b^i\b^j\,\frac{C_{ij}}{2E_\Phi} \right\rangle (t_1)
 +O(\ve^3).
\end{eqnarray*}
The combinatorics is the same as before; we thus obtain
\begin{eqnarray*}
\frac{\Phi^*[t_1,t_2,t_3,t_4]}{[t_1,t_2,t_3,t_4]}-1&=&
 \frac{1}{4}\left\langle \ve
 X,\a\frac{E_\Phi'}{E_\Phi}+\b^i\frac{D_iE_\Phi}{E_\Phi}\right\rangle^2(t_1)\\[6pt]
&+\displaystyle{\half}&\left\langle \ve X\otimes \ve
 X,\a^2\left(\frac{E_\Phi'}{E_\Phi}\right)'
+2\a\b^i\left(\frac{D_iE_\Phi}{E_\Phi}\right)'+\b^i\b^j
 D_j\left(\frac{D_iE_\Phi}{E_\Phi}\right)\right\rangle(t_1)\\[6pt]
&-2&\left\langle \ve X\otimes \ve X,\a^2\,\frac{A}{6E_\Phi}+\a\b^i\,
 \frac{B_i}{2E_\Phi}+\b^i\b^j\,\frac{C_{ij}}{2E_\Phi}\right\rangle(t_1)+O(\ve^3).
\end{eqnarray*}
As in the case $N=1$, see (\ref{devt_cr_AB}) and (\ref{devt-cr}), we still have $A=E_\Phi''-\psi'\cdot\psi''$ and also $B_i=\frac{1}{2}D_i E_\Phi'+\psi'\cdot D_i\psi'$. We now collect the terms according to
\begin{eqnarray}\label{ABC}
\nonumber
\frac{\Phi^*[t_1,t_2,t_3,t_4]}{[t_1,t_2,t_3,t_4]}-1&=&\left\langle \ve
 X\otimes\ve X, \a^2 
\left(\frac{1}{6}\frac{E_\Phi''}{E_\Phi}+\frac{\psi'\cdot\psi''}{3E_\Phi}-\frac{1}{4}\left(\frac{E_\Phi'}{E_\Phi}\right)^2\right)\right\rangle(t_1)\\[6pt]\nonumber
&&+\left\langle \ve X\otimes\ve X, 
\a\b^i\left(\frac{1}{2}\frac{D_iE_\Phi'}{E_\Phi}-\frac{\psi'\cdot D_i\psi'}{E_\Phi}-\frac{E_\Phi'D_iE_\Phi}{2E_\Phi^2}\right)\right\rangle(t_1)\\[6pt]\nonumber
&&+\left\langle \ve X\otimes\ve X,\b^1\b^2\left(\frac{D_2
 D_1E_\Phi}{E_\Phi}-\frac{2C_{12}}{E_\Phi}-\frac{D_2E_\Phi D_1E_\Phi}{2
 E_\Phi^2}\right)\right\rangle(t_1)\\[6pt]
&&+O(\ve^3).
\end{eqnarray}
We denote by $S(\Phi)$ the coefficient of $\a^2$, $S_i(\Phi)$ that of $\a\b^i$ and
$S_{12}(\Phi)$ that of $\b^1\b^2$. Let us start with the computation of $S_{12}(\Phi)$, our goal being to write $C_{12}$ as a function of $\rE_\Phi$ and its derivatives. 

We first give a useful lemma.

\begin{lem}\label{bonD1D2}
For any $\phi,\widetilde{\phi}\in\cC^\infty(\rS^{1|2})^2$, the following relations hold:
\begin{equation}
(\phi\cdot D_1\psi)(\widetilde{\phi}\cdot D_1\psi)+(\phi\cdot
 D_2\psi)(\widetilde{\phi}\cdot D_2\psi)=\phi\cdot\widetilde{\phi} \;E_\Phi,
\label{ClosureFormula}
\end{equation}
and also,
\begin{equation}\label{timesdot}
\phi\times D_2\psi=\l \phi\cdot D_1\psi
\qquad \text{and}\qquad \phi\times D_1\psi=-\l \phi\cdot D_2\psi,
\end{equation}
with $\l^2=1$, 
%depending only on $\psi$, 
and where the cross-product is defined by
$\phi\times\widetilde{\phi}=\phi_1\widetilde{\phi}_2-\phi_2\widetilde{\phi}_1$.\\
Moreover, $\psi'$ being odd, for even $\phi$ and $\widetilde{\phi}$, we have
\begin{equation}\label{dottimes}
(\psi'\cdot\phi)(\psi'\cdot\widetilde{\phi})=\psi'_1\psi'_2(\phi\times\widetilde{\phi}).
\end{equation}
\end{lem}
\begin{proof}
Proposition \ref{Dortho} proves the first equality. As $D_2\psi\cdot
 D_2\psi=E_\Phi$, either $D_2\psi_1$ or $D_2\psi_2$ is invertible, where $\psi=(\psi_1,\psi_2)$. Suppose that $D_2\psi_2$ is invertible, we then have $D_1\psi_1=\l D_2\psi_2$ for
 some $\l$. Using $D_i\psi\cdot D_j\psi=\delta_{ij}E_\Phi$, we obtain
 $D_1\psi_2=-\l D_2\psi_1$ and $\l^2=1$. If $D_2\psi_1$ is invertible the
 same equalities hold. Then, easy calculation ends the proof.
\end{proof}

We have $C_{12}=D_2 D_1\varphi'+\psi\cdot D_2 D_1\psi'$, as given by Lemma \ref{N,taylordevt}. 
Differentiating the constraint $D_i\varphi=\psi\cdot D_i\psi$, see (\ref{N,contactconstraint}), we find
 $D_1\varphi'=\psi'\cdot D_1\psi+\psi\cdot D_1\psi'$, and then $D_2
 D_1\varphi'=D_2\psi'\cdot D_1\psi-\psi'\cdot D_2 D_1\psi+D_2 \psi\cdot
 D_1\psi'-\psi\cdot D_2 D_1\psi'$. Plugging the latter expression into
 $C_{12}$, and using $D_i\psi\cdot D_j\psi=0$, for $i\neq j$, we obtain
 $C_{12}=-\psi'\cdot D_2 D_1\psi$. 
Using the proof of Lemma \ref{bonD1D2}, we have $D_1\psi_1=\l D_2 \psi_2$ and $D_1\psi_2=-\l D_2 \psi_1$, and then  $\psi'\cdot D_2 D_1\psi= 2\l\psi'_1\psi'_2$. Moreover, as $\frac{1}{4}D_1E_\Phi D_2E_\Phi=(\psi'\cdot D_1\psi)(\psi'\cdot D_2\psi)$, we find, using
 (\ref{dottimes}) and (\ref{timesdot}), $\frac{1}{4}D_1E_\Phi
 D_2E_\Phi=\l\psi'_1\psi'_2 E_\Phi$. We thus have $C_{12}=-\frac{1}{2E_\Phi}D_1E_\Phi
 D_2E_\Phi$, and replacing this in the last expression of $S_{12}$, as
 given by (\ref{ABC}), we finally get
\begin{equation}\label{S12}  
S_{12}(\Phi)=\frac{D_2 D_1E_\Phi}{E_\Phi}-\frac{3}{2}\frac{D_2E_\Phi
 D_1E_\Phi}{ E_\Phi^2}.
\end{equation}
We will show that $S_1=\frac{1}{2}D_2 S_{12}$, and then, exchanging
 $D_1$ and $D_2$, readily obtain $S_2=-\half D_1 S_{12}$. Let us first
recall the expression of $S_1$, given in (\ref{ABC}),
$$
S_1(\Phi)=\frac{1}{2}\left(\frac{D_1 E_\Phi'}{E_\Phi}-\frac{E_\Phi'D_1
 E_\Phi}{E_\Phi^2}-\frac{2\psi'\cdot D_1\psi'}{E_\Phi}\right).
$$
Secondly, we find
$$
D_2 S_{12}(\Phi)=\frac{ D_1E_\Phi'}{E_\Phi}-\frac{D_2 E_\Phi D_2
 D_1E_\Phi}{E_\Phi^2}+\frac{3}{2}\frac{D_2E_\Phi D_2D_1E_\Phi}{
 E_\Phi^2}-\frac{3}{2}\frac{E_\Phi' D_1E_\Phi}{ E_\Phi^2}.
$$
We then have to show that the following expression vanishes, namely
\begin{equation}\nonumber
2S_1(\Phi)-D_2 S_{12}(\Phi)=-\frac{2\psi'\cdot
 D_1\psi'}{E_\Phi}+\frac{1}{2}\frac{D_2E_\Phi D_1D_2E_\Phi}{ E_\Phi^2}+\frac{1}{2}\frac{E_\Phi' D_1E_\Phi}{
 E_\Phi^2}.
\end{equation}
To that end, let us use Formula (\ref{N,E}) to rewrite the last two
 terms as $D_2E_\Phi D_1D_2E_\Phi=4(\psi'\cdot D_2\psi)(D_1\psi'\cdot
 D_2\psi+\psi'\cdot D_2D_1\psi)$, and $E_\Phi' D_1E_\Phi= 4(\psi'\cdot
 D_1\psi)(D_1\psi'\cdot D_1\psi)$, respectively. We have already proved that
 $\psi'\cdot D_2 D_1\psi= 2\l\psi'_1\psi'_2$, and using
 (\ref{ClosureFormula}), we thus obtain
\begin{equation}\label{S1}
2S_1-D_2 S_{12}=0.
\end{equation}

At last, we want to show that $6S=D_1D_2 S_{12}+\frac{1}{2}S_{12}^2$.
 Begin by writing explicitly
$$
6S(\Phi)=\frac{E_\Phi''}{E_\Phi}-\frac{3}{2}\left(\frac{E_\Phi'}{E_\Phi}\right)^2+\frac{2\psi'\cdot\psi''}{E_\Phi},
$$
with the help of \ref{ABC}, and also
\begin{eqnarray*}
D_1D_2 S_{12}(\Phi)
&=&
\frac{E_\Phi''}{E_\Phi}-\frac{D_1E_\Phi
 D_1E_\Phi'}{E_\Phi^2}-\frac{3}{2}\left(\frac{E_\Phi'}{E_\Phi}\right)^2
-\frac{3}{2}\frac{D_1E_\Phi' D_1E_\Phi}{E_\Phi^2}\\[6pt]
&&+\frac{1}{2}\frac{D_2E_\Phi D_2E_\Phi'}{ E_\Phi^2}%\\[6pt]%\nonumber
-\frac{1}{2}\left(\frac{D_2D_1E_\Phi}{ E_\Phi}\right)^2-\frac{D_1E_\Phi
 D_2E_\Phi D_2D_1E_\Phi}{ E_\Phi^3}.
\end{eqnarray*}
We now compute the difference:
\begin{eqnarray*}
D_1D_2 S_{12}(\Phi)-6S(\Phi)
&=&
\frac{1}{2}\frac{D_1E_\Phi D_1E_\Phi'+D_2E_\Phi D_2E_\Phi'}{ E_\Phi^2}
 -\frac{D_1E_\Phi D_2E_\Phi D_2D_1E_\Phi}{ E_\Phi^3}\\
&&-\frac{1}{2}\left(\frac{D_2D_1E_\Phi}{
 E_\Phi}\right)^2-2\frac{\psi'\cdot\psi''}{E_\Phi}.
\end{eqnarray*}
Using Equation (\ref{N,E}), we get $\frac{1}{2}(D_1E_\Phi
 D_1E_\Phi'+D_2E_\Phi D_2E_\Phi')=2[(\psi'\cdot D_1\psi)(\psi''\cdot
 D_1\psi)+(1\leftrightarrow 2)]+2[(\psi'\cdot D_1\psi)(\psi'\cdot
 D_1\psi')+(1\leftrightarrow 2)]$. Thanks to Formula (\ref{ClosureFormula}), the first term
 reduces to $2\psi'\cdot \psi'' E_\Phi$, and using (\ref{dottimes}) the second
one turns out to be $2\psi'_1\psi'_2(D_1\psi\times D_1\psi'+D_2\psi\times D_2\psi')$, which is equal to $4\l\psi'_1\psi'_2(D_2\psi\cdot D_1\psi')$, in view of (\ref{timesdot}). On the other hand, using the previous equalities, we find $D_1E_\Phi D_2E_\Phi D_2D_1E_\Phi=-8\l E_\Phi\psi'_1\psi'_2(D_1\psi'\cdot D_2\psi)$; hence,
 we obtain
$$
D_1D_2 S_{12}(\Phi)-6S(\Phi)= -\frac{3}{2}\frac{D_1E_\Phi D_2E_\Phi D_2D_1E_\Phi}{
 E_\Phi^3}-\frac{1}{2}\left(\frac{D_2D_1E_\Phi}{ E_\Phi}\right)^2,
$$
which is the desired result: $6S=D_1D_2 S_{12}+\frac{1}{2}S_{12}^2$. Together with (\ref{S1}) and (\ref{S12}), the latter equation finishes the derivation of (\ref{2Sch}). Using the projection (\ref{aD1D2}) on $1$-densities, we obtain the Schwarzian derivative $\rS$, given by (\ref{2RSch}).

An argument analogous to that of Remark \ref{rmkprojection} helps us to prove that the cocycles~$\rS$ and $\cS$ are nontrivial.

To find the kernel of the cocycle $\rS$, whence that of $\cS$, we
 have to use another form of $\rS$, namely
$$\rS(\Phi)=-2E_\Phi^\half(D_2D_1(E_\Phi^{-\half}))\a.$$
Hence, if $\Phi\in\ker(\rS)$, we have $D_2D_1(E_\Phi^{-\half})=0$. As was the case for $N=1$, the solution is $E_\Phi=(cx+d+\d\cdot\xi)^{-2}$. A direct computation shows that there exists
 $h\in\PC(2|2)$ such that $E_\Phi=E_{\widehat{h}}$. Since the kernel of
the Euclidean cocycle is $\widehat{\rE}_+{1|N}/\{\pm\Id\}$ for~$N=2$, we obtain as announced: $\ker(\rS)=\PC(2|2)$.  
The proof of Theorem \ref{thm2Sch} is complete.
\end{proof}

%%%%%%%%%%%%%%%%%%%%%%%%%%%%%%%%%%%%%%%%%%%%%%%%%%%%%%%%%%%%
%%%%%%%%%%%%%%%%%%%%%%%%%%%%%%%%%%%%%%%%%%%%%%%%%%%%%%%%%%%%
\section{Conclusion, discussion and outlook}
\label{Conclusion}
%%%%%%%%%%%%%%%%%%%%%%%%%%%%%%%%%%%%%%%%%%%%%%%%%%%%%%%%%%%%
%%%%%%%%%%%%%%%%%%%%%%%%%%%%%%%%%%%%%%%%%%%%%%%%%%%%%%%%%%%%
Starting off with the orthosymplectic group, $\SpO(2|N)$, and two nested
 subgroups $\rE_+(1|N)\subset\Aff_+(1|N)\subset\SpO(2|N)$, we have been able to uniquely characterize Euclidean, $I_\mathsf{e}$, affine, $I_\mathsf{a}$, and projective, $I_\mathsf{p}$, invariants for their actions as contactomorphisms of the
 supercircle $\rS^{1\vert N}$, using the central notion of $p|q$-transitivity. Moreover, these invariants are characteristic of
 their defining groups. For $N=0,1,2$, their even part does
 characterize the image of the supergroups $\EO(1|N)$, $\AO(1|N)$ and $\SpO(2|N)$, by the projective action on $\rS^{1|N}$, within the group, $K(N)$, of all contactomorphisms. In doing
 so, we have recovered in a systematic fashion the previously
 introduced \cite{Aok,Nel} even and odd cross-ratios. 
 
 Then, using a natural super
 extension of the Cartan formula (\ref{TheCartanFormula}), we have
 provided a novel construction of the nontrivial $1$-cocycles $\cE$, $\cA$, and $\cS$ of $K(N)$, associated with the even
 invariants, for $N=0,1,2$. We have also succeeded to recover the known expressions \cite{Fri,Coh,Rad0,Rad} of the
 Schwarzian derivatives for $N=0,1,2$. The kernels of the above-mentioned
 $1$-cocycles have been shown to coincide with the groups defining the
 invariants leading to them. So, for each geometry, the group action, the even invariant and the $1$-cocycle are three equivalent geometric objects on the super\-circle~$\rS^{1|N}$ (where $N=0,1,2$), endowed with its standard contact structure. 
 
 \goodbreak
 
 In the cases $N=0,1$, a complete classification of the subgeometries of the contact geometry of $\rS^{1|N}$ is given by that of the nontrivial cohomology spaces $H^1(K(N),\cF_\l)$, see Theorem \ref{ThmH1} and Remark \ref{rmkN=1}. A similar classification for $N=2$ is still lacking. Work in progress related to the determination of $H^1(K(2),\cF_\l)$ should provide a first insight into this classification, as well as that of the $1$-cohomology spaces of $K(2)$ with coefficients in other natural modules such as $\Omega^1(\rS^{1|2})$ and $\cQ(\rS^{1|2})$. In doing so, we will resort to the computation of $H^1(k(2),\cF_\l)$ carried out by Ben Fraj \cite{BFr}.
 
For $N>2$, our method yields, indeed, the Euclidean and affine
 $1$-cocycles of $K(N)$. There is, however, no way to obtain, in our approach, Radul's
 Schwarzian integro-differential operator for $N=3$, since there exists
 no projection from $\cQ(\rS^{1\vert 3})$ to $\cF_\half=k(3)_\mathrm{reg}^*$ intertwining
 the $K(3)$ action. Moreover, our study provides a clear cut explanation of the fact that $\cS(\Phi)$ cannot be derived as a quadratic differential by the Cartan formula (see Remark \ref{N=3_obstruction}) for $N\geq 3$, and therefore help us understand why the Radul expression for $N=3$ involves pseudo-differential operators. 
 
 \goodbreak
      
We have, so far, studied the supercircle $\rS^{1|N}$; but there are in
 fact two super\-extensions of the circle, namely $\rS^{1|N}$ and
 $\rS_+^{1|N}$, see \cite{FL,Rad,GLS}. Let us discuss the case~$N=1$. The
 only difference between these two supermanifolds is that the functions on $\rS^{1|1}$ are, indeed, functions on
 $\bbR^{1|1}$ invariant with respect to the transformation
 $(x,\xi)\mapsto(x+2\pi,\xi)$, whereas functions on the M\"obius supercircle, $\rS_+^{1|1}$, can be viewed as functions on $\bbR^{1|1}$ invariant under the transformation $(x,\xi)\mapsto(x+2\pi,-\xi)$. Here the coordinate $x$ is regarded as an angular coordinate on $\rS^1$. The canonical contact structure on $\bbR^{1|1}$ define a contact structure
 on both $\rS^{1|1}$ and $\rS_+^{1|1}$ \cite{Ovs}.
 All our cocycles, prior to projections, are left invariant by the map
 $(x,\xi)\mapsto(x,-\xi)$, as well as the projections themselves; then $\cE$,
 $\cA$, $\cS$, and $\rA$, $\rS$ still define cocycles on $\rS_+^{1|1}$. This can be generalized for $N>1$ along the same line as before. 

We expect that our approach will help us express the Bott-Thurston 
cocycles of $K(1)$ and $K(2)$ given by Radul in terms of the Berezin integral of the cup product of the $1$-cocycles $\cE$ and $\cA$ introduced above (in a manner similar to the case of $\Diff_+(\rS^1)$ spelled out in \cite{GR}). This should, hence, extend the classical formula worked out in \cite{DG} using a contact $1$-form on $\Diff_+(\rS^1)\times\bbR$. 

Another plausible development would be the superization of the hyperboloid of one sheet in $\mathrm{sl}(2,\bbR)^*$ whose conformal geometry is related to the projective geometry of null infinity \cite{KS,DG}.

%%%%%%%%%%%%%%%%%%%%%%%%%%%%%%%%%%%%%%%%%%%%%%%%%%%%%%%%%%%%
%%%%%%%%%%%%%%%%%%%%%%%%%%%%%%%%%%%%%%%%%%%%%%%%%%%%%%%%%%%%

\end{document}